\documentclass[lineno]{jfm_clean} 
\usepackage{newtxtext}
\usepackage{newtxmath}

\usepackage[inter-unit-product = \ensuremath{{}\cdot{}}]{siunitx}

\usepackage{bm} 

\usepackage{multirow}
\usepackage{booktabs}

\newcommand{\RomanNumeralCaps}[1]
\linenumbers

\usepackage[labelformat=simple]{subcaption}
\newcommand*{\defeq}{\stackrel{\text{def}}{=}}
\newcommand{\llangle}{\langle\!\langle}
\newcommand{\rrangle}{\rangle\!\rangle}
\newcommand{\rd}{\mathrm{d}}
\newcommand{\ri}{\mathrm{i}}
\newcommand{\re}{\mathrm{e}}

\newcommand\Wor{{\text{Wo}}}  
\newcommand\Str{{\text{St}}}  

\usepackage{natbib}
\usepackage[doipre={doi:}]{uri}
\DeclareUrlCommand\doi{\def\UrlLeft##1\UrlRight{\href{http://dx.doi.org/##1}{doi:##1}}\urlstyle{rm}}

\usepackage{hyperref}
\hypersetup{
    colorlinks = true,
    urlcolor   = blue,
    linkcolor  = blue,
    citecolor  = blue,
    hypertexnames = false,
}

\title{Oscillatory flows in three-dimensional deformable microchannels}

\shorttitle{Oscillatory flows in deformable microchannels}

\author{Anxu Huang\aff{1}\footnote{These authors contributed equally.}, Shrihari D.\ Pande\aff{2}$\dagger$, Jie Feng\aff{1}\corresp{Corresponding authors: {jiefeng@illinois.edu}, {christov@purdue.edu}.} \and Ivan C.\ Christov\aff{2}$\ddagger$}

\shortauthor{Huang, Pande, Feng \& Christov}

\affiliation{\aff{1}Department of Mechanical Science and Engineering, University of Illinois at Urbana-Champaign, Urbana, Illinois 61801, USA
\aff{2}School of Mechanical Engineering, Purdue University, West Lafayette, Indiana 47907, USA}

\begin{document}
\maketitle

\begin{abstract}
Deformable microchannels emulate a key characteristic of soft biological systems and flexible engineering devices: the flow-induced deformation of the conduit due to slow viscous flow within. Elucidating the two-way coupling between oscillatory flow and deformation of a three-dimensional (3D) rectangular channel is crucial for designing lab- and organ-on-a-chip microsystems and eventually understanding flow-structure instabilities that can enhance mixing and transport. To this end, we determine the axial variations of the primary flow, pressure, and deformation for Newtonian fluids in the canonical geometry of a slender (long) and shallow (wide) 3D rectangular channel with a deformable top wall under the assumption of weak compliance and without restriction on the oscillation frequency (\textit{i.e.}, on the Womersley number). Unlike rigid conduits, the pressure distribution is not linear with the axial coordinate. To validate this prediction, we design a PDMS-based experimental platform with a speaker-based flow-generation apparatus and a pressure acquisition system with multiple ports along the axial length of the channel. The experimental measurements show good agreement with the predicted pressure profiles across a wide range of the key dimensionless quantities: the Womersley number, the compliance number, and the elastoviscous number. Finally, we explore how the nonlinear flow--deformation coupling leads to self-induced streaming (rectification of the oscillatory flow). Following Zhang and Rallabandi (\textit{J.\ Fluid Mech.}, vol.~996, 2024, A16), we develop a theory for the cycle-averaged pressure based on the primary problem's solution, and we validate the predictions for the axial distribution of the streaming pressure against the experimental measurements.
\end{abstract}



\section{Introduction}
\label{sec:intro}

Fluid-structure interactions (FSIs) between oscillatory internal viscous fluid flows and their deformable confining boundaries are ubiquitous across natural and engineered systems and across scientific disciplines. For example, such FSIs arise in biomedical problems involving blood circulation \citep{P80, F97} in the cardiovascular system \citep{vosse2011pulse,menon2024cardiovascular}, specifically the large arteries \citep{K97,GJ04,CLMT05}, as well as flows in the vocal cords \citep{HH11}, lungs \citep{G94,GJ04,HH11}, brain \citep{Gan23,BLCJNB23}, retina \citep{SF19} and synovial joints \citep{dowson1986micro,PBG22}. Harnessing these FSIs has proven critical for the design and construction of microfluidic devices \citep{LESKUBL09,XWZZW21,BWW22,BPCOJ22,Muduetal24,mosadegh2010integrated}, which has enabled emerging technologies such as organs-on-chips \citep{BI14,Lindetal17,DBG22,Leungetal22}, flexible and wearable electronics \citep{kwon2023battery,cheol2018wearable}, and soft robotics \citep{EG14,MEG17,BSBGO18,matia2023harnessing,xu2023compact}. The deformation of compliant conduits by oscillatory flows also arises in elastohydrodynamic lubrication \citep{KCC21,R24} and at scales relevant to geophysical problems \citep{KSSR16,RCG23}. While oscillatory (and/or pulsatile) viscous flows \citep{Z00} in elastic tubes is a time-honored subject, dating back to Womersley's work in the 1950s \citep{W55,W55a}, the two-way-coupled interplay between the pressure gradient driving the flow and the deformation of the compliant wall has, surprisingly, not been fully explored, as a series of recent works highlighted \citep{PWC23,ZR24,KB25}.

The initial impetus for understanding the fluid mechanics of oscillatory flow was that ``the central problem in haemodynamics flow ... [was not] satisfactorily resolved for arterial flow'' \citep{mcdonald1955relation}. This ``central problem,'' initially defined by \citet{burton1952laws} as ``the relation of pressure to flow,'' has been a cornerstone in studies focusing on the flow rate--pressure drop ($q - \Delta p$) relationship of oscillatory internal flows, starting with the fundamental study by \citet{W55}. However, despite significant progress, much of the literature has focused on the case of one-way coupling between flow and deformation, namely how the flow and structure behave if the pressure gradient is considered known \textit{a priori}, as assumed by \citet{W55a,W57}. Furthermore, prior experimental investigations of such internal periodic flows in compliant conduits generally focused on biomedical applications. For example, \citet{pielhop2012analysis, pielhop2015experimental} and \citet{dorner2021experimental} used ``time-resolved particle-image velocimetry combined with a wall detection algorithm and non-invasive pressure measurements'' to study the biofluid mechanics of large elastic PDMS vessels under moderate-to-high Reynolds number conditions, including for a non-Newtonian blood-analog fluid. However, these studies lacked theoretical support to rationalize their observations. Most recently, \citet{KB25} performed detailed simulations of the two-way-coupled FSI between an oscillatory flow and a thin viscoelastic shell, with the analysis and interpretation guided by Womersley's classical theory.

Returning to the microfluidic context, the resistance of channels of various cross-sectional shapes is well understood \citep{bruus2008theoretical}. Recently, significant progress has been made in developing the $q - \Delta p$ relationship for two-way-coupled \emph{steady} flows through deformable conduits at low Reynolds number \citep{C21}, the same is not true for similar oscillatory flows. As \citet{DDS20} note, ``pulsatile microfluidics is still in its infancy,'' especially when it comes to wall compliance. In the context of low-Reynolds-number flows, \citet{AC20} and \citet{PWC23} revisited the problem of two-way-coupled FSI between oscillatory flows and elastic channels and tubes, showing (through asymptotic analysis, modeling, and direct numerical simulations) the existence of a secondary (streaming) flow resulting from the nonlinear coupling between pressure and deformation at low Reynolds number. This streaming effect represents a type of \emph{self-induced} peristaltic pumping mechanism, which is a topic extensively studied in biomechanics \citep[see, \textit{e.g.},][and the references therein]{TB11,ACB23}. The opposite problem of an external flow driven by oscillations of an elastic body, termed \emph{soft streaming}, was considered by \citet{BPG22} and \citet{CBG24}. \citet{L58} and \citet{LA72} argued that, for flows in arteries, both the advective nonlinearity of the Navier--Stokes equations and the geometric and material nonlinearity of the elastic wall should both be taken into account. \citet{LA72} provided limited comparisons between experimental measurements of velocity, flow rate, and shear stress and simulations via a reduced-order model as support. Indeed, the complete theory by \citet{ZR24} of the \emph{elastoinertial rectification} mechanism, underlying the streaming flow observed by \citet{PWC23}, shows that advective inertia is inextricably coupled to pressure and deformation in these flows, but using linear elastic theories [\textit{e.g.}, thin shell theory in \citep{ZR24}],  suffices in the context of microfluidic flows. However, the nonzero cycle-averaged pressure (due to the nonlinear coupling between flow inertia and wall deformation) has not been systematically measured in an experiment.

The desire to fill this knowledge gap in the field of \emph{nonlinear microfluidics} \citep{XWZZW21,BWW22} motivates the present combined theoretical--experimental study. Specifically, we develop a microfluidic experimental platform consisting of a test section with a compliant wall and multiple pressure ports. To drive the flow, we develop a custom pressure-generation system capable of delivering a wide range of frequencies and amplitudes with precise control. This experimental platform enables us to characterize the spatiotemporal pressure distribution due to the oscillatory flow of Newtonian fluids in 3D deformable microchannels, including measuring the weak cycle-averaged (streaming) pressure. To guide and rationalize the experiments, we extend the axisymmetric theory of \citet{ZR24} to three-dimensional (3D) slender and shallow deformable channels, which are commonly encountered in experimental microfluidic systems \citep{GEGJ06,CTS12,OYE13,MY19,PDMJ24}.

The rest of this paper is organized as follows: In \S~\ref{sec:3D_wide}, we introduce the problem, governing equations, scales, and dimensionless numbers and apply the lubrication approximation. In \S~\ref{sec:exp}, we describe the experimental setup for the oscillatory flow generation and methodology for pressure measurement. In \S~\ref{sec:ei_recti}, we perform a perturbation expansion for weak compliance to obtain the primary flow, pressure, and wall displacement profiles for a Newtonian fluid in such a slender and shallow deformable 3D channel, as well as the streaming pressure distribution along the channel. We cross-validate our results through quantitative comparisons between experimental pressure measurements and theoretical predictions in \S~\ref{sec:discussion}. Conclusions and perspectives for future work are summarized in \S~\ref{sec:conclusion}.


\section{Oscillatory flow in a slender and shallow 3D deformable channel}
\label{sec:3D_wide}

\begin{figure}
    \centering
    \includegraphics[width=0.85\textwidth]{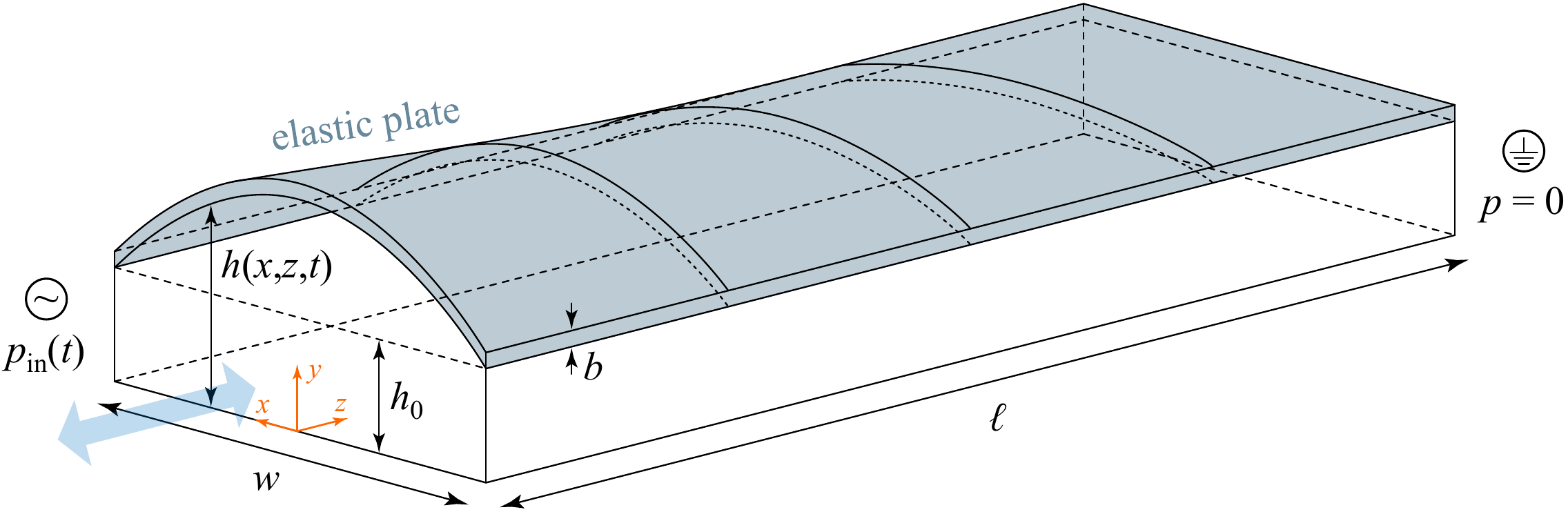}
    \caption{Schematic of the 3D deformable shallow and slender rectangular microchannel geometry of initial (undeformed) height $h_0$, axial length $\ell$, and transverse width $w$, such that $\ell \gg w \gg h_0$. The top wall (darker color) is an elastic plate structure of thickness $b$ that can deform from $y=h_0$ to $y=h(x,z,t)$, where $h(x,z,t)-h_0=u_y(x,z,t)$ is the vertical displacement of the fluid--solid interface. The top wall is clamped (no displacement) along the planes $x=\pm w/2$ (and $0\le z\le \ell$), while taking the outlet pressure as gauge, $p|_{z=\ell}=0$, ensures no deformation along the plane $z=\ell$ (and $-w/2 \le x \le +w/2$). An oscillatory inlet pressure, $p|_{z=0} = p_\mathrm{in}(t)$ of amplitude $p_0$ and angular frequency $\omega$, drives the flow.}
    \label{schematic}
\end{figure}

Consider the oscillatory flow of a Newtonian fluid with constant density $\rho_f$ and constant dynamic viscosity $\mu_f$ through a 3D rectangular channel of initial height $h_0$, transverse width $w$, and axial length $\ell$, shown schematically in figure~\ref{schematic}, as commonly encountered in experimental microfluidic systems. The flow is driven by inlet pressure oscillations of magnitude $p_0$ and angular frequency $\omega$. The velocity field is $\bm{\varv} = (\varv_x,\varv_y,\varv_z)$, the pressure field is $p$. The bottom ($y=0$) and side  ($x=\pm w/2$) walls are rigid, but the top wall can deform. The deformable top wall is made from a linearly elastic material with Young's modulus $E$ and Poisson's ratio $\nu_s$. The displacement field of the top wall is given by $\bm{u} = (u_x,u_y,u_z)$. 

\subsection{Governing equations of the oscillatory flow}

In this context, \citet{MCSPS19} analyzed the inertialess start-up flow, following onto the steady problem solved by \citet{CCSS17}, introducing scalings that balance all velocity components in the conservation of mass equation. \citet{RAB21} also used these scalings for a non-Newtonian version of the problem. Here, unlike these prior works, we adopt the scaling of the 3D problem introduced by \citet{BC22}, which leads to a leading-order problem that can be spanwise-averaged. Although both approaches are valid within the assumed order of approximation for a shallow and wide channel, only the latter approach allows us to make analytical progress in the oscillatory flow problem. Specifically, we scale \emph{both} cross-sectional velocities, $\varv_x$ and $\varv_y$, so that they are smaller than the axial one, $\varv_z$, by a factor of $\epsilon \defeq h_0/\ell$, where $\epsilon \ll1$ for a slender channel. 

With all this in mind, we introduce the dimensionless variables for the problem:
\begin{multline}
    X = \frac{x}{w},\qquad Y = \frac{y}{h_0},\qquad Z = \frac{z}{\ell},\qquad T = \frac{t}{\omega^{-1}},\qquad V_X = \frac{\varv_x}{\epsilon \varv_c},\qquad V_Y = \frac{\varv_y}{\epsilon \varv_c},\\
    V_Z = \frac{\varv_z}{\varv_c}, \qquad P = \frac{p}{p_0}, \qquad U_X = \frac{u_x}{u_c},\qquad U_Y = \frac{u_y}{u_c},\qquad U_Z = \frac{u_z}{u_c}, \qquad H = \frac{h}{h_0}.
    \label{eq:nd_vars}
\end{multline}
Here, $p_0$ and $\omega$ are the amplitude and angular frequency, respectively, of the inlet pressure oscillations driving the flow. The scales $\varv_c$ and $u_c$ for the axial velocity and wall displacement, respectively, will be determined below. 
In terms of the dimensionless variables from \eqref{eq:nd_vars}, the governing partial differential equations (PDEs) of leading-order-in-$\epsilon$ unsteady flow in a 3D channel [\textit{i.e.}, a slender channel under the lubrication approximation for which $\epsilon \ll 1$ \citep{L07}] are
\begin{subequations}\begin{align}
    \delta \frac{\partial V_X}{\partial X} + \frac{\partial V_Y}{\partial Y} + \frac{\partial V_Z}{\partial Z} &= 0, \label{eq:com_3D}\\
    0 &= - \frac{\partial P}{\partial X},\\
    0 &= - \frac{\partial P}{\partial Y},\\
    \Wor^2 \left[ \frac{\partial V_Z}{\partial T} + \frac{\beta}{\gamma} \left( \delta V_X \frac{\partial V_Z}{\partial X} + V_Y \frac{\partial V_Z}{\partial Y} + V_Z \frac{\partial V_Z}{\partial Z} \right) \right] &= - \frac{\partial P}{\partial Z} + \delta^2 \frac{\partial^2 V_Z}{\partial X^2} + \frac{\partial^2 V_Z}{\partial Y^2}, \label{eq:zmom_3D}    
\end{align}\label{eq:lubrication_pdes_3D}\end{subequations}
where we have chosen the axial velocity scale as $\varv_c = h_0^2 p_0/(\ell \mu_f)$ to balance viscous and pressure forces in \eqref{eq:zmom_3D}. In scaling the problem this way, several key dimensionless numbers emerge:
\begin{subequations}\begin{alignat}{2}
    \delta &\defeq \frac{h_0}{w},\\
    \Wor^2 &\defeq \frac{h_0^2/(\mu_f/\rho_f)}{\omega^{-1}} &&= \frac{\text{transverse momentum diffusion timescale}}{\text{oscillation timescale}},\displaybreak[3]\\
    \beta &\defeq \frac{u_c}{h_0} &&= \frac{\text{top wall displacement scale}}{\text{initial channel height}},\\
    \gamma &\defeq \frac{u_c/(\epsilon \varv_c)}{\omega^{-1}} &&= \frac{\text{elastoviscous timescale}}{\text{oscillation timescale}}.
\end{alignat}\label{eq:key_dimless_groups}\end{subequations}
Here, $\delta$ is the channel's cross-sectional (inverse) aspect ratio ($\delta \ll 1$ for wide channels); $\beta$ is the compliance number \citep{CCSS17}, where wall displacement scale $u_c$ will be introduced below upon specifying the elasticity model; $\Wor$ is the Womersley number \citep{W55a}; $\gamma$ is the elastoviscous number \citep{ZR24,EG14}, where the elastoviscous time scale emerges from comparing the vertical displacement scale $u_c$ (set by elasticity) to the vertical velocity scale $\epsilon \varv_c$ (set by the balance of viscous and pressure forces). For a stiff top channel wall (weakly compliant channel), we would expect $\beta\ll1$. On the other hand, we do not impose any restrictions on $\gamma$ or $\Wor$ \textit{a priori}. Although it may be tempting to rewrite $\beta/\gamma$ in \eqref{eq:zmom_3D} as a Reynolds number \citep{PWC23}, it will become clear below why it is not a good idea \citep{ZR24}.

Now, assuming a wide (shallow) channel, $\delta\ll1$, so that we can neglect all terms at $O(\delta)$ and higher in \eqref{eq:lubrication_pdes_3D}, we have:
\begin{subequations}\begin{align}
    \frac{\partial V_Y}{\partial Y} + \frac{\partial V_Z}{\partial Z} &= 0, \label{eq:com_3D_wide}\\
    0 &= - \frac{\partial P}{\partial X}, \label{eq:xmom_3D_wide}\\
    0 &= - \frac{\partial P}{\partial Y}, \label{eq:ymom_3D_wide}\\
    \Wor^2 \left[ \frac{\partial V_Z}{\partial T} + \frac{\beta}{\gamma} \left( V_Y \frac{\partial V_Z}{\partial Y} + V_Z \frac{\partial V_Z}{\partial Z} \right) \right] &= - \frac{\partial P}{\partial Z} + \frac{\partial^2 V_Z}{\partial Y^2}. \label{eq:zmom_3D_wide}    
\end{align}\end{subequations}
As usual, from \eqref{eq:xmom_3D_wide}  and \eqref{eq:ymom_3D_wide}, we immediately conclude that $P=P(Z,T)$ only.

The flow obeys no-slip and no-penetration conditions along the bottom wall of the channel:
\refstepcounter{equation}
$$
    V_Z|_{Y=0} = 0,\qquad V_Y|_{Y=0} = 0.
    \eqno{(\theequation{\mathit{a},\mathit{b}})}
    \label{eq:no_slip_bc}
$$%
Along the deformable (thus, moving) top wall, $y=h(x,z,t)$, a kinematic condition, $\bm{\varv} = D\bm{u}/Dt$, applies. In dimensionless component form,  the kinematic condition is:
\begin{subequations}\begin{align}
    V_X|_{Y=H} &= \left.\left(\gamma \frac{\partial U_X}{\partial T} + \frac{u_c}{w} V_X\frac{\partial U_X}{\partial X} + \frac{u_c}{h_0} V_Y\frac{\partial U_X}{\partial Y} + 0\right)\right|_{Y=H} ,\label{eq:kinematic_bc_x}\\ 
    V_Y|_{Y=H} &= \left.\left(\gamma \frac{\partial U_Y}{\partial T} + \frac{u_c}{w} V_X \frac{\partial U_Y}{\partial X} + \frac{u_c}{h_0} V_Y \frac{\partial U_Y}{\partial Y} + 0\right)\right|_{Y=H},\label{eq:kinematic_bc_y}\\ 
    V_Z|_{Y=H} &= O(\epsilon) ,\label{eq:kinematic_bc_z}
\end{align}\label{eq:kinematic_bc_all}\end{subequations}%
having used the third condition to simplify the first two. Further simplification can only be obtained upon specifying the governing equations of the wall deformation (in \S~\ref{sec:deformation} below).
Since we restrict ourselves to pressure-driven flows (as will be described in more detail in \S~\ref{sec:exp} below), the inlet and outlet pressures are known:
\refstepcounter{equation}
$$
    P|_{Z=0} = P_\mathrm{in}(T),\qquad P|_{Z=1} = 0.
    \eqno{(\theequation{\mathit{a},\mathit{b}})}
    \label{eq:pressure_bc}
$$

\subsection{Governing equations of the top wall deformation}
\label{sec:deformation} 

Assuming a thin structure, we shall employ \emph{plate} theory. The central tenet of plate theory is that the structure is thin and thus the vertical displacement $u_y$ (in the thin direction) does not vary with $y$ \citep{reddy07}. Thus, the bottom surface of the top wall (\textit{i.e.}, the fluid--solid interface) is located at $h_0+u_y$ upon deformation of the wall (see figure~\ref{schematic}).

\citet{AMC20} provides a full derivation and discussion of the equations of motion of slender and shallow \emph{thick} plates obeying the theory due to \citet{R45} and \citet{M51}, which governs the flexural deformation of isotropic, elastic plates deduced from the three-dimensional equations of elasticity when the plate's thickness is nontrivial---a so-called first-order shear deformation theory \citep{reddy07,CE19}. Here, we summarize the key details. The in-plane plane displacements are given by $u_x = u_x^0(x,z) + \bar{y} \phi_x(x,z)$ and $u_z = u_z^0(x,z) + \bar{y} \phi_z(x,z)$, where $\phi_x$ and $\phi_z$ are rotations of the normal vector to the plate's midsurface about the $x$- and $z$-axis, respectively, and $\bar{y}$ is measured from the midsurface, \textit{i.e.}, $\bar{y} = y - h_0 - b/2$. The PDEs for $u_x^0(x,z)$ and $u_z^0(x,z)$ decouple from those for $\phi_y(x,z)$, $\phi_z(x,z)$, and $u_y(x,z)$. In the absence of external in-plane loads, these PDEs are homogeneous. Thus, from the clamped boundary conditions (BCs) (as in figure~\ref{schematic}), it follows that $u_x^0 = u_z^0 \equiv 0$. Consequently, the plate problem reduces to determining the vertical displacement, $u_y$, and the rotations of the normal, $\phi_x$ and $\phi_z$. A balancing argument shows that the appropriate scales for $\phi_x$ and $\phi_z$ are $u_c/w$ and $u_c/\ell$, respectively.

For a slender and shallow plate, neglecting terms of $O(\delta)$ and $O(\epsilon)$, $\Phi_Z$ drops out of the governing plate equations, which now feature only $U_Y$ and $\Phi_X$ \citep{AMC20}. Our starting point is these equations, at the leading order in $\delta$ and $\epsilon$, namely 
\begin{subequations}\begin{align}
    \Str_s\frac{\partial^2 U_Y}{\partial T^2} &= \frac{1}{720}\frac{\partial^3 \Phi_X}{\partial X^3} + P(Z,T), \label{eq:bend_mindlin1_2}\\
    0 &= \Phi_X + \frac{\partial U_Y}{\partial X} - \frac{\mathscr{T}}{60} \frac{\partial^2\Phi_X}{\partial X^2},
    \label{eq:bend_mindlin2_2}
\end{align}\label{eq:thick_plate_eqs_2}%
\end{subequations}
where $\mathscr{T} \defeq 10(b/w)^2/(1-\nu_s)$ is a dimensionless parameter quantifying the shallowness of the plate. Having balanced transverse bending with the pressure load from the flow in \eqref{eq:bend_mindlin1_2}, $u_c = w^4 p_0 / (720 D_b)$, where $D_b = E b^3/[12(1-\nu_s^2)]$ is the flexural rigidity of a plate in pure bending \citep{reddy07}, $E$ is Young's modulus, and $\nu_s$ is Poisson's ratio. The factor of $1/720$ included in $u_c$ might not be obvious now; it is included to eliminate all numerical prefactors in \eqref{eq:continuity_3D_3} below. 

Similar to \citet{MCSPS19}, we have retained the displacement's inertia in \eqref{eq:bend_mindlin1_2}, but neglected rotary inertia in \eqref{eq:bend_mindlin2_2} as is standard in the literature \cite[see, \textit{e.g.},][Chap.~13]{ZTZ13}. The inertia of the plate is thus quantified by the solid's Strouhal number 
\begin{equation}
    \Str_s \defeq \frac{\rho_s b u_c/p_0}{\omega^{-2}} = \frac{(\text{solid's timescale for response to loading})^2}{(\text{oscillation timescale})^2}, 
\end{equation}
where $\rho_s$ is the density of the elastic wall material.

Generalizing the steady results of \citet{SC18}, equations~\eqref{eq:thick_plate_eqs_2} describe \emph{transverse} bending of the wall, \textit{i.e.}, the deformation in any $(X,Y)$ plane for fixed $Z$, subject to the normal load imposed by the hydrodynamic pressure $P(Z,T)$. The \emph{axial} tension and bending are neglected as they scale with $\epsilon\ll1$ \citep{AMC20}. 
The corresponding clamped BCs supplementing \eqref{eq:thick_plate_eqs_2} are
\refstepcounter{equation}
$$
    U_Y |_{X=\pm 1/2} = 0,\qquad \Phi_X |_{X=\pm 1/2} = 0.
    \eqno{(\theequation{\mathit{a},\mathit{b}})}
    \label{eq:clamped_bc_RM}
$$    

Observe that if we let $\mathscr{T}\to0$ (\textit{i.e.}, $b/w\to0$), then $\Phi_X = - \partial U_Y/\partial X$ from \eqref{eq:bend_mindlin2_2}, and substituting this relation into \eqref{eq:bend_mindlin1_2} and (\ref{eq:clamped_bc_RM}\textit{b}) reduces these to the corresponding equation of Kirchhoff--Love (thin-plate) theory \citep[see, \textit{e.g.},][]{HKO09}.

Finally, based on this discussion, we have deduced that $U_X = O(\delta)$ and $U_Z = O(\epsilon)$, thus the in-plane displacements are \emph{negligible} compared to the vertical (out-of-plane) displacement $U_Y$ at the leading order, which significantly simplifies the kinematic condition~\eqref{eq:kinematic_bc_all}.

\subsection{Coupling flow to deformation}

Since the in-plane displacements $U_X$ and $U_Z$ are negligible, the dimensionless kinematic condition~\eqref{eq:kinematic_bc_all} reduces to three independent BCs. The $X$- and $Z$-components, \eqref{eq:kinematic_bc_x} and \eqref{eq:kinematic_bc_y}, are essentially no-slip BCs: $V_X|_{Y=H}=V_Z|_{Y=H}=0$. While the $Y$-component \eqref{eq:kinematic_bc_y} dictates that the fluid's vertical velocity matches that of the wall:
\begin{equation}
    V_Y|_{Y=H} = \frac{\gamma}{\beta} \frac{\partial H}{\partial T},
    \label{eq:kinematic_bc}
\end{equation}
where $H(X,Z) = 1 + \beta U_Y(X,Z)$ is the dimensionless height of the fluid channel. We recognize that $\gamma/\beta$ can also be interpreted as a fluidic Strouhal number \citep{R83,WW19,IWC20,PWC23}. Equation \eqref{eq:kinematic_bc} is key to the two-way coupling of the flow and deformation.

Next, we seek to manipulate the kinematic condition \eqref{eq:kinematic_bc} into one involving the flow rate. To this end, we define the flow rate $Q$ by integrating over the cross-sectional area in an  $(X,Y)$ plane,
\begin{equation}
    Q = \int_{-1/2}^{+1/2} \int_0^H V_Z \,\rd Y \rd X.
    \label{eq:flow_rate_3D}
\end{equation}
Then, a lengthy but straightforward calculation using the Leibniz rule to swap the order of $Y$-integration and $Z$-differentiation shows that the kinematic BC~\eqref{eq:kinematic_bc} can be combined with \eqref{eq:flow_rate_3D} to re-express the conservation of mass equation~\eqref{eq:com_3D_wide} as the \emph{continuity equation}: 
\begin{equation}
    \frac{\partial Q}{\partial Z} 
    + \frac{\gamma}{\beta} \frac{\partial}{\partial T} \int_{-1/2}^{+1/2} \big[ 1 + \beta U_Y(X,Z,T) \big] \,\rd X = 0.
    \label{eq:continuity_3D_2}
\end{equation}

\subsection{Summary of the governing equations and dimensionless numbers}

In summary, \eqref{eq:zmom_3D_wide}, \eqref{eq:thick_plate_eqs_2}, and \eqref{eq:continuity_3D_2} are the governing equations of the oscillatory flow in a 3D shallow and slender channel with a plate-like deformable top wall in pure bending.

\begin{table}
    \begin{center}
    \def~{\hphantom{0}}
    \begin{tabular}{l@{\qquad} l@{\qquad} l@{\qquad} l}
    Dimensionles number & Notation/expression & Assumption & Typical value/range \\[5pt]

    Channel's slenderness & $\epsilon=h_0/\ell$ & negligible & $0.008$ \\[2pt]

    Channel's shallowness & $\delta = h_0/w$ & negligible & $0.08$ \\[2pt]
      
    Womersley & $\Wor=h_0\sqrt{\rho_f \omega/\mu_f}$ & none & $0.5-3.2$ \\[2pt]
    
    Elastoviscous & $\gamma = w^4 \ell^2 \mu_f \omega/(720 D_b h_0^3)$ & none & $0.15-15$ \\[2pt]
  
    Compliance & $\beta = w^4p_0/(720 D_b h_0)$ & small & $0.01-0.2$ \\[2pt]
    
    Solid's inertial Strouhal & $\Str_s = {\rho_s b w^4 \omega^2}/(720 D_b)$ & negligible & $1.7\times 10^{-6} - 1.4\times 10^{-3}$ \\[2pt]

    Plate's shallowness & $\mathscr{T} = 10(b/w)^2/(1-\nu_s)$ & small & $0.03-0.138$
    \end{tabular} 
    \caption{Key dimensionless numbers of the 3D elastoinertial rectification problem, based on the characteristic displacement scale $u_c$ for a plate and a characteristic axial velocity scale $\varv_c$ under lubrication theory. Typical values/ranges are based on the experimental setup (\S~\ref{sec:exp}). Negligible numbers are taken as zero in the analysis (\textit{i.e.}, the theory is a ``at leading order'' in these parameters), while small quantities are taken into account; perturbatively in the case of $\beta$.}
    \label{tab:DimlessNum} 
    \end{center}
\end{table}

The key dimensionless groups of the problem are summarized in table~\ref{tab:DimlessNum}. The experimental system, to be discussed next in \S~\ref{sec:exp}, was designed to achieve an oscillatory flow with $\Wor,\gamma=O(1)$ in the weakly compliant regime $\beta$ small (but not negligible) in a slender $\epsilon\ll1$ and shallow $\delta\ll1$ channel, which are the key approximations made in the theoretical analysis. These considerations lead to the parameter values/ranges given in table~\ref{tab:DimlessNum}. Additionally, we learn from table~\ref{tab:DimlessNum} that the channels constructed have negligible wall inertia, $\Str_s\ll1$, allowing us to consider the plate deformation as quasi-static.


\section{Experimental setup}\label{sec:exp}
\subsection{Oscillatory flow generation and shaping}\label{art}

Our experimental setup is shown in figure~\ref{expstp}. Previous experimental investigations of oscillatory flow generation in microchannels \citep{vishwanathan2020generation, vishwanathan2022assembly,vishwanathan2023synchronous,levenstein2022millimeter, pielhop2015experimental,dorner2021experimental,RDC19} involved synchronizing mechanical vibrations directly to the fluid itself in a microfluidic channel. Notably, the way that \citet{vishwanathan2020generation, vishwanathan2022assembly, vishwanathan2023synchronous} achieved this is by attaching an inlet tube to the diaphragm of a speaker, such that the fluid within the channel was driven by the oscillations initiated from the inlet tube, ultimately creating an oscillatory flow. Inspired by this approach, we customized our oscillatory flow generation module as follows. We utilized a function generator (GH-CJDS66, Koolertron) connected with a speaker (DR-US200275, Drok) to ensure a robust signal input. We further introduced a polydimethylsiloxane (PDMS, Sylgard 184, Dow Corning) liquid chamber positioned towards the speaker, with its membrane linked to the speaker diaphragm via a rigid, 3D-printed connector to facilitate efficient mechanical vibration transmission. The chamber was first fabricated using 3D-printed molding techniques, and the membrane (with thickness on the order of 0.5 mm) sealing the liquid chamber was then bonded by inverting the chamber onto a liquid layer of PDMS mixture, which was subsequently cured at $90\si{\celsius}$ for one hour to ensure solidification and secure adhesion. All PDMS components were fabricated with a 10:1 (w/w) ratio of silicone elastomer base to curing agent. 

In each experiment, this chamber was filled with the working fluid to generate sufficient pressure amplitude. Once the analog sinusoidal signal from the function generator was transmitted into the speaker, which enabled its diaphragm to vibrate, the connected deformable membrane of the chamber thereby vibrated, causing the fluid oscillations inside the chamber and transmitting them to the microchannel. Before each experiment, we carefully eliminated any entrapment of air within the entire interior space, such as air pockets in all pressure ports and tubing, which is found to be important for obtaining reliable pressure amplitudes by the pressure transducer. In addition, the microchannel outlet was submerged in a liquid reservoir filled with the working fluid slightly above the level of the microchannel to maintain a constant hydrostatic pressure at the outlet, which is approximately the atmospheric pressure given the small height.

\subsection{Fabrication of the microchannels with deformable top walls}\label{channeldescription}

\begin{figure}
    \centering
    \includegraphics[width=\textwidth]{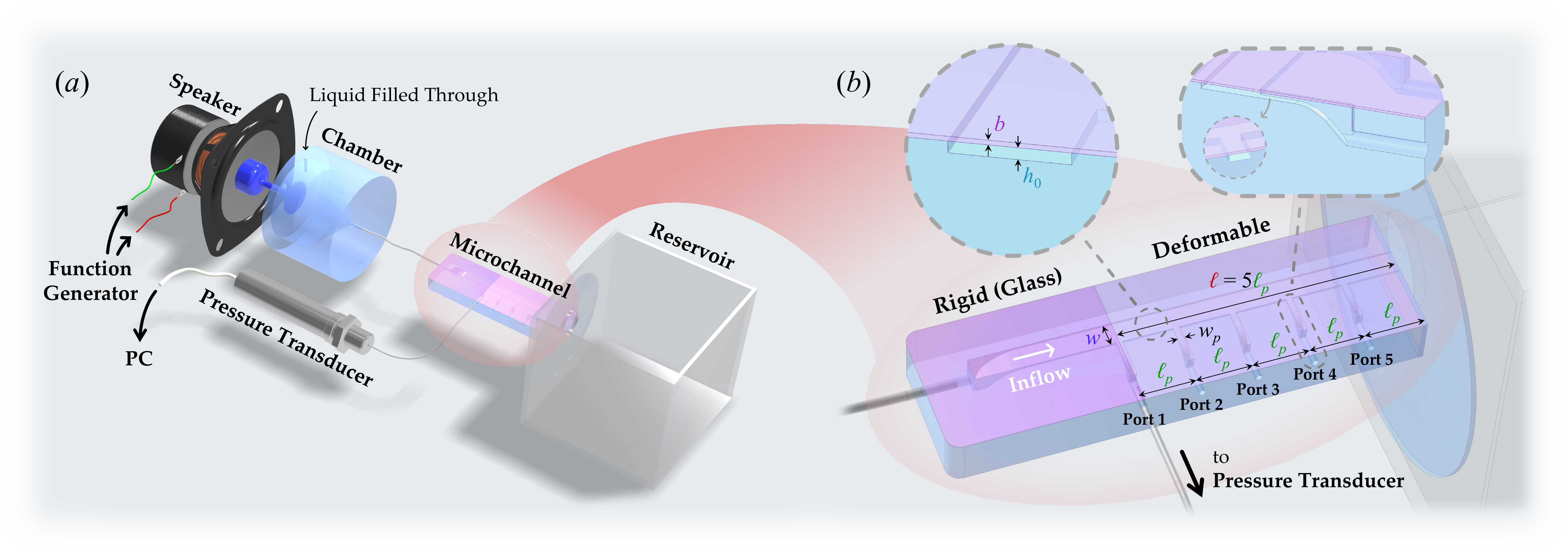}
    \caption{Experimental system with oscillatory flow in a 3D deformable rectangular microchannel. (\textit{a}) Setup schematic. The entire interior space of the system is completely filled with the fluid prior to the experiments. To initiate the flow, an analog sinusoidal signal generated by the function generator is transmitted into the speaker, enabling its diaphragm to vibrate. The deformable membrane of the liquid chamber (linked to the speaker diaphragm via a rigid, 3D-printed connector shown in dark blue) transmits these vibrations, causing the oscillation of the fluid within both the chamber and the following channel. (\textit{b}) Microchannel configuration. The channel features five pressure ports connecting to the data acquisition system (pressure transducer \& PC). The five ports (each of width $w_p$) are evenly spaced with an axis-to-axis interval $\ell_p$ in the flow direction. The microchannel section between the first port and the outlet is covered with a deformable PDMS film of length $\ell$ at the top, and the section of the channel ahead of the first port is covered by a rigid glass slide, with its front edge precisely aligned to the center of the first port.}
    \label{expstp}
\end{figure}

To fabricate the rectangular microchannel with deformable top walls, we follow the same procedure as \citet{CBCF24}, where a 3D-printing technique (Mars Resin 3D Printer, ELEGOO, USA) was employed to manufacture the reverse mold with the designated channel dimensions as listed in table~\ref{tab:channel}. A mixing ratio of 10:1 (w/w) between the silicone elastomer base and the curing agent was used to prepare the PDMS elastomer. The PDMS mixture was subsequently poured into the 3D-printed mold and degassed under vacuum for 1 hour to fully remove entrapped air bubbles. The mixture was then cured in an oven at $90\si{\celsius}$ for 24 hours. Upon curing, the PDMS block as the channel substrate was carefully detached from the mold, followed by the fluid inlet and pressure measurement ports being precisely punctured by a 2-mm disposable biopsy punch. To probe the time-varying pressure distribution along the flow direction, the five ports [spaced $\ell_p$ apart, see figure~\ref{expstp}(\textit{b})] were designed to connect to a pressure transducer for pressure measurements along the channel. All ports (with width $w_p$ at the branching point where they intersect with the main channel) were located to the side of the channel [figure~\ref{expstp}(\textit{b})] for convenient experimental operation. The lateral cross-sectional view of the entire port is shown in the inset of figure~\ref{expstp}(\textit{b}). Additionally, a smooth geometric transition following an elliptic arc was employed at the microchannel inlet to minimize any secondary flows created at sharp corners.

\begin{table}
    \begin{center}
    \def~{\hphantom{0}}
    \begin{tabular}{c@{\quad}c@{\quad}c@{\quad}c@{\quad}c}
        \multicolumn{5}{c}{Microchannel}\\[5pt]
        $h_0$ (\si{\milli\meter})&
        $w$ (\si{\milli\meter})&
        $\ell$ (\si{\milli\meter})&
        $\ell_p$ (\si{\milli\meter})&
        $w_p$ (\si{\milli\meter})\\[2pt]
        \multirow{2}{*}{\(0.50 \pm 0.005\)} & 
        \multirow{2}{*}{\(5.0 \pm 0.05\)} & 
        \multirow{2}{*}{\(60.0 \pm 0.25\)} & 
        \multirow{2}{*}{\(12.0 \pm 0.05\)} & 
        \multirow{2}{*}{\(1.4 \pm 0.05\)}
    \end{tabular}\\[15pt] 
    \begin{tabular}{c@{\qquad}c@{\qquad}c@{\qquad}c}
        \multicolumn{4}{c}{Deformable top walls}\\[5pt]
        $b$ (\si{\milli\meter}) & 
        $E$ (\si{\mega\pascal}) & 
        $\nu_s$ (--) & 
        $\rho_s$ (\si{\kilo\gram/\meter\cubed})\\[2pt]
        \(0.43 \pm 0.01\) & \(1.02 \pm 0.05\) & \multirow{2}{*}{\(0.47 \pm 0.1\)} & \multirow{2}{*}{\(1070 \pm 10\)}\\
        \(0.20 \pm 0.005\) & \(1.21 \pm 0.02\)  & & 
    \end{tabular}
    \caption{Dimensions of the microchannel and elastic properties of the deformable walls.}
    \label{tab:channel}
    \end{center}
\end{table}

Two thin PDMS films were used as the deformable top walls. One PDMS film with a thickness of $0.43 \pm 0.01~\si{\milli\meter}$ was fabricated in the lab following the procedure of \citet{CBCF24}. The other PDMS film (GASKET-UT-200PK) with a thickness of $0.2 \pm 0.005~\si{\milli\meter}$ was purchased from SIMPore Inc., USA. The films and microchannel substrate were treated with a 4.5 MHz hand-held corona treater (BD-20AC, Electro-Technic Products, USA) for 30 seconds, and then we brought the film and the channel substrate together into conformal contact for bonding. We note that this bonding approach for oscillatory flows was proven to be robust, as confirmed by the reproducibility of our experiments. We further attached a rigid glass slide on top of the thin PDMS film, other than the section between the first port and the outlet, to confine the deformability solely to this section. We confirmed all dimensions of the channel and the thin PDMS film by microscope visualization (table~\ref{tab:channel}). We also measured the Young modulus $E$ and the Poisson ratio $\nu_s$ of the thin PDMS film using Dynamic Mechanical Analysis (Q800 DMA, TA instruments, USA) at a room temperature of $20\si{\celsius}$. For the PDMS films that we fabricated in the lab, the measurements of the Young moduli were performed after 24 hours to ensure the material properties of the PDMS films were fully stabilized. The density $\rho_s$ from the datasheet is reported.

To validate our experimental system for oscillatory flows, we also fabricated a rigid microchannel, which is the counterpart of the deformable microchannels with the same dimensions and configuration, except that the top wall was replaced by a large PDMS block. Using the same fabrication procedures as for the channel substrate, the PDMS block was fabricated to be as thick as the channel substrate (thickness $\approx1~\si{\centi\meter}$), ensuring that its rigidity prevents any deformation under the typical imposed hydrodynamic pressure (amplitude $\approx0.1~\si{\kilo\pascal}$) used for the experiments with the thin-film top walls. Subsequently, the large PDMS block was attached to its dedicated channel substrate by undergoing the same surface treatment as for the thin-film top walls described above.

\subsection{Working fluids and pressure measurements}

Two different viscous Newtonian fluids, deionized (DI) water and an aqueous solution of 50 wt\% glycerin, were used to tune the Womersley number $\Wor$ and the elastoviscous number $\gamma$. The working fluids' densities and viscosities are reported in table~\ref{tab:fluids}, which were measured right before each experiment. The viscosity of the working fluids was quantified by flow sweep tests performed on a stress-controlled rheometer (DHR-3, TA Instruments, USA). 

For the pressure measurement, the gauge pressure at each pressure port was recorded by a pressure transducer (PX409-10WGUSBH, OMEGA, USA) wired to the PC with a control software (Digital Transducer Application, OMEGA, USA), using a sampling rate of $1,000~\si{\hertz}$ for all cases, which is much larger than the frequency of the pressure signal in our experiments (typically set between $1$ and $16~\si{\hertz}$). Prior to each experiment, the pressure transducer was calibrated using a column of water. Then, the entire interior space of the system, including the tubing connected to the pressure transducer, was completely filled with the working fluid. Thereafter, the top opening of the PDMS chamber (used to fill it with fluid) was kept closed, and the channel outlet was completely submerged under the free surface of the fluid in the reservoir. The five pressure ports were also blocked by the tubing connected to the pressure transducer and four plugs to maintain a hermetically sealed liquid environment.

\begin{table}
    \begin{center}
    \def~{\hphantom{0}}
    \begin{tabular}{c@{\qquad} c@{\qquad} c}
        Fluid & $\rho_f$ (\si{\kilogram/\cubic\meter}) & $\mu_f$ (\si{\milli\pascal~\second}) \\[5pt]
        Deionized (DI) water & $1000\pm1$ & $1 \pm 0.05$ \\
        50 wt\% glycerin solution & $1115\pm1.1$ & $6\pm0.3$
    \end{tabular}
    \caption{Physical properties of the working fluids.}
    \label{tab:fluids}
    \end{center}
\end{table}

We measured the pressure distribution of both the primary flow (derived in \S~\ref{sec:primary_flow}) and the secondary streaming flow (derived in \S~\ref{sec:secondary_flow}). For the primary flow, we synchronized the pressure's time variation in each port with one pressure transducer as follows. At the onset of each experiment, the function generator was configured to produce a sinusoidal waveform with a predetermined frequency and voltage. As the speaker diaphragm continuously oscillated under the excitation of the function generator, pressure variations over time were recorded. For a given pressure input, characterized by an amplitude and frequency, we first recorded data at the first port, $p_1(t)=p(z_1,t)$, as shown in figure~\ref{expstp}(\textit{b}) (hereafter referred to as ``port 1'') for a sufficiently long duration. Next, to synchronize the $p_i(t)$ data of port $i$ (where $i = 2$, 3, 4, or 5) onto the time axis of port 1's data under the same pressure input, we implemented a staged recording method for port $i$. First, we began the data collection on port 1. Second, we transferred the pressure sampling tube from port 1 to port $i$. Third, we continued data collection at port $i$ before concluding the session. This entire process was completed within the total duration of port 1's data acquisition in the first session. To align the pressure datasets, we applied a temporal adjustment by shifting the whole $p_i(t)$ curve of the second session along the time axis until its initial time slot measuring port 1 precisely overlapped with port 1's reference data from the first session, eliminating any phase discrepancy. This approach ensured that the data recorded for each port was accurately synchronized with the time axis established by port 1's dataset. The same procedure was systematically repeated for all pressure measurement ports, ultimately yielding a fully synchronized pressure dataset across all five ports for a given pressure input. 

On the other hand, to measure the weak streaming pressure in the secondary flow for each pressure port, we first obtained a baseline pressure value with no flow (the speaker off), then recorded the pressure signal with the flow (the speaker on). The difference between the cycle-averaged value of the pressure signal with the flow and the baseline pressure reflects the streaming effect therein. We used these pressure differences to further calculate the cycle-averaged streaming pressure, which, in this case, were the sole measurements we are ultimately interested in, meaning the synchronization between ports was not critical.


\section{Theory of elastoinertial rectification in a 3D deformable channel}
\label{sec:ei_recti}

The oscillatory flow has a time-harmonic (zero-mean) component called the \emph{primary} flow. Nonlinearity, due to two-way coupling between the flow and deformation as well as flow inertia, induces a \emph{secondary} flow. This secondary flow can have a nonzero mean component, which is often referred to as a \emph{streaming} flow. Classically, streaming is studied in the context of viscous flows in which weak advective inertia provides the nonlinearity. The latter example is referred to as \emph{acoustic streaming} \citep{Riley98,R01,S12}. Either way, if a zero-mean forcing produces a nonzero net cycle-averaged flow, we say that the oscillatory flow experiences \emph{rectification}.

For internal flows in conduits, a nonzero mean flow can be produced by a slowly varying conduit cross-section (\textit{e.g.}, the radius in a tube) or by curvature (twists and turns) of the conduit in the flow direction \citep{P80}. Streaming flows can also be generated by peristalsis (small-amplitude waves propagated along the conduit wall), with applications to microelectromechanical systems (MEMS) \citep{SS01} and biofluid mechanics \citep{RSGG20,TPZR25}. In the present context of an internal flow in a deformable channel, the streaming flow arises from the combination of geometric (flow-induced elastic deformation) and inertial nonlinearities. 

To decouple the primary and secondary flows, we follow \citet{ZR24} and expand all variables in a perturbation series in $\beta\ll1$,
\begin{equation}
    \{ V_Y, V_Z, Q, P \} = \underbrace{\{ V_{Y,0}, V_{Z,0}, Q_0, P_0  \}}_{\text{primary flow}} \; + \; \beta \underbrace{\{ V_{Y,1}, V_{Z,1}, Q_1, P_1 \}}_{\text{secondary flow}} \; + \; O(\beta^2),
    \label{eq:beta_expand}
\end{equation}
without restricting $\gamma$ or $\Wor$. Below, we construct the primary flow solution explicitly (\S~\ref{sec:primary_flow}), while for the secondary flow, we only evaluate the cycle-averaged (\textit{i.e.}, streaming or rectified) component (\S~\ref{sec:secondary_flow}).

\subsection{Primary flow: \texorpdfstring{$O(1)$}{O(1)} solution}
\label{sec:primary_flow}

Substituting the expansion \eqref{eq:beta_expand} into the momentum equations~\eqref{eq:xmom_3D_wide} and \eqref{eq:ymom_3D_wide}, we find that the primary pressure $P_0$ does not vary with $X$ or $Y$. Then, at $O(1)$, the $Z$-momentum equation~\eqref{eq:zmom_3D_wide} and the no-slip BCs from (\ref{eq:no_slip_bc}\textit{a}) and (\ref{eq:kinematic_bc_all}\textit{c}) become
\begin{equation}
\left. \begin{array}{l}
    \displaystyle
    \Wor^2  \frac{\partial V_{Z,0}}{\partial T} = - \frac{\rd P_0}{\rd Z} + \frac{\partial^2 V_{Z,0}}{\partial Y^2},  \\[16pt]
    \displaystyle
    V_{Z,0}|_{Y=0} = V_{Z,0}|_{Y=1} = 0.
\end{array} \right\}
\label{eq:Vz0_prob}
\end{equation}
We expect the primary flow to be harmonic, thus we introduce \emph{phasors}: $V_{Z,0}(Y,Z,T)=\Real[V_{Z,0,a}(Y,Z)\re^{\ri T}]$ and $P_{0}(Z,T)=\Real[P_{0,a}(Z)\re^{\ri T}]$. Substituting the phasors into \eqref{eq:Vz0_prob}, the solution for the axial velocity phasor's amplitude is easily found \citep[see, \textit{e.g.},][]{PWC23} to be 
\begin{equation}
    V_{Z,0,a}(Y,Z) = 
    \frac{1}{\ri{\Wor}^2}\left[1-\frac{\cos\left(\ri^{3/2} (1-2Y){\Wor}/2\right)}{\cos\left(\ri^{3/2} {\Wor}/2 \right)}\right] \left( -\frac{\rd P_{0,a}}{\rd Z} \right).
    \label{eq:Vza0_soln}
\end{equation}
Now, the flow rate phasor's amplitude is evaluated by substituting \eqref{eq:Vza0_soln} into \eqref{eq:flow_rate_3D} to yield
\begin{equation}
\left. \begin{array}{l}
    \displaystyle
    Q_{0,a} = \int_{-1/2}^{+1/2} \int_{0}^{1} V_{Z,0,a} \,\rd Y \rd X = \mathfrak{f}(\Wor) \left(-\frac{\rd P_{0,a}}{\rd Z}\right),\\[16pt]
    \displaystyle
    \mathfrak{f}(\Wor) \defeq \frac{1}{\ri \Wor^2}\left[1 - \frac{1}{\ri^{3/2}{\Wor}/2}\tan\left(\ri^{3/2}{\Wor}/2\right)\right].
\end{array} \right\}    
\label{eq:Qa0_soln}
\end{equation}

Next, we determine the top wall displacement. Having verified that $\Str_s\ll1$, we neglect the wall's inertia. Then, the solution of \eqref{eq:thick_plate_eqs_2} subject to the BCs \eqref{eq:clamped_bc_RM} \citep[see, \textit{e.g.},][]{SC18} is 
\begin{equation}
    U_Y(X,Z,T) = 30 \left(\frac{1}{4} - X^2\right) \left[ \left(\frac{1}{4} - X^2\right) + \frac{\mathscr{T}}{5} \right] P(Z,T).
    \label{eq:beam_3D_soln}
\end{equation}
Substituting \eqref{eq:beam_3D_soln} into  \eqref{eq:continuity_3D_2}, performing the $X$ integration, and then substituting the perturbation expansions for $Q$ and $P$ in terms of phasors, the continuity equation becomes
\begin{equation}
    \frac{\rd Q_{0,a}}{\rd Z} + \left(1 + \mathscr{T}\right) \gamma \ri P_{0,a}(Z) = 0.
    \label{eq:continuity_3D_3}
\end{equation}
Finally, substituting the flow rate--pressure gradient relation~\eqref{eq:Qa0_soln}, into the continuity equation~\eqref{eq:continuity_3D_3}, and taking into account the pressure BCs~\eqref{eq:pressure_bc}, we arrive at a boundary-value problem (BVP) for the primary pressure amplitude:
\begin{equation}
\left. \begin{array}{l}
    \displaystyle
    \mathfrak{f}(\Wor) \frac{\rd^2 P_{0,a}}{\rd Z^2} = 
    \ri \left(1 + \mathscr{T}\right) \gamma P_{0,a}(Z), \\[16pt]
    \displaystyle
    P_{0,a}(0) = 1,\quad 
    P_{0,a}(1) = 0.
\end{array} \right\}
\label{eq:P0_BVP}
\end{equation}
Here, according to the experimental system's setup, we have taken $P_\mathrm{in}(T) = \Real[\re^{\ri T}] + O(\beta)$, having imposed the amplitude of the pressure oscillations as the characteristic pressure scale via \eqref{eq:nd_vars}. We have left open the possibility of $O(\beta)$ corrections to the oscillatory pressure BC, which may arise from how the oscillatory flow was generated in the experiments. We return to this issue in \S~\ref{sec:secondary_flow} below.

\begin{figure}
    \centering
    \begin{subfigure}[h]{0.45\textwidth}
        \caption{}
        \includegraphics[width=\textwidth]{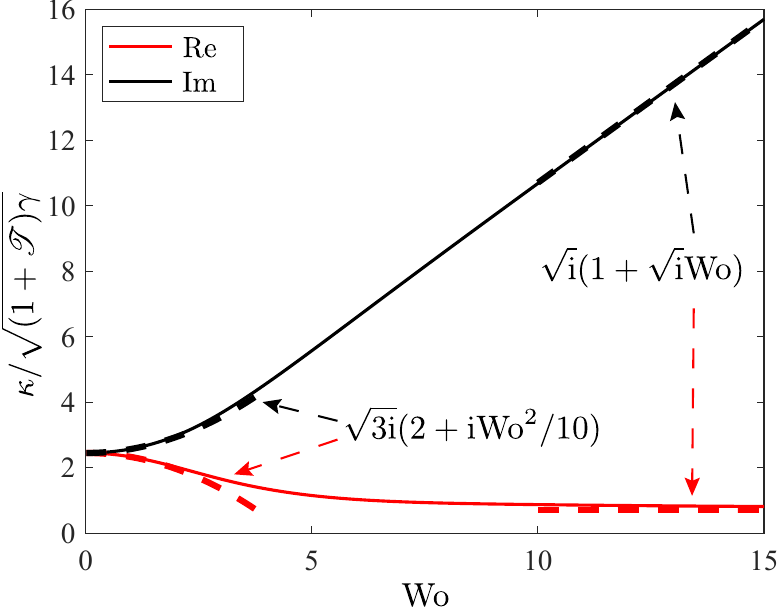}
    \end{subfigure}
    \quad
    \begin{subfigure}[h]{0.45\textwidth}
        \caption{}
        \includegraphics[width=\textwidth]{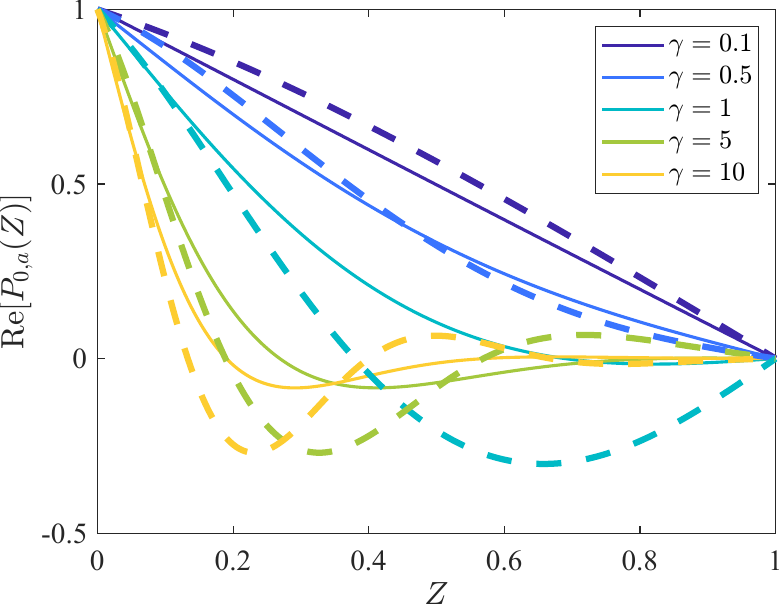}
    \end{subfigure}
    \caption{(a) Dependence of the reduced complex ``wavenumber'' $\kappa/\sqrt{(1+\mathscr{T})\gamma} = \sqrt{\ri/\mathfrak{f}(\Wor)}$ on $\Wor$ and its asymptotic behaviors (dashed curves labeled with arrows). (b) Shape of the primary pressure amplitude's axial distribution, $\Real[P_{0,a}(Z)]$ from (\ref{eq:Pa0_soln}\textit{a}), for $\Wor = 1$ (solid) and $\Wor = 3$ (dashed) and across a range of $\gamma$ values.}
    \label{fig:kappa_P0a_Z}
\end{figure}

The solution to the BVP~\eqref{eq:P0_BVP} is easily found to be
\refstepcounter{equation}
$$
    P_{0,a}(Z) = \frac{\sinh\big(\kappa(1-Z)\big)}{\sinh \kappa},\qquad 
    \kappa = \kappa(\Wor,\gamma,\mathscr{T}) \defeq 
    \sqrt{\frac{\ri\left(1 + \mathscr{T}\right)\gamma}{\mathfrak{f}(\Wor)}},    \eqno{(\theequation{\mathit{a},\mathit{b}})}
    \label{eq:Pa0_soln}
$$%
which has the same form as the corresponding solution in an axisymmetric deformable tube \citep{ZR24,DG91,R83}. This solution is illustrated in figure~\ref{fig:kappa_P0a_Z}. We observe that $\kappa/\sqrt{(1 + \mathscr{T})\gamma}$ is solely a function of $\Wor$ with asymptotics of $\sim \sqrt{3\ri} (2 + \ri\Wor^2/10)$ as $\Wor\to0$ and $\sim \sqrt{\ri} (1 + \sqrt{\ri} \Wor)$ as $\Wor\to\infty$. The large-$\Wor$ asymptotics show a much faster ($\sim\Wor$) growth for the 3D rectangular channel compared to the 3D axisymmetric tube ($\sim\sqrt{\Wor}$) in \citep{ZR24}.

Observe that as the elastoviscous adjustment of the wall becomes instantaneous compared to the oscillation time scale ($\gamma\to0$), \textit{i.e.}, the channel is effectively \emph{rigid}, then $\kappa\to0$ and $P_{0,a}(Z) \to 1 - Z$ from \eqref{eq:Pa0_soln}, as expected.


\subsection{Secondary flow: \texorpdfstring{$O(\beta)$}{O(beta)} solution}
\label{sec:secondary_flow}

To define the secondary (streaming) flow problem, we must determine the axial velocity at the deformable wall in terms of known quantities. Following, \citet{BSC22}, we proceed by domain perturbation \citep{L82,L07}. To this end, recall that $H=1+\beta U_Y = 1+\beta U_{Y,0}+O(\beta^2)$, and expand the axial velocity at the wall in a Taylor series:
\begin{equation}
    V_Z|_{Y=H} =  V_{Z,0} |_{Y=1} + \left. \beta U_{Y,0} \frac{\partial V_{Z,0}}{\partial Y}\right|_{Y=1} + \beta V_{Z,1}|_{Y=1} + O(\beta^2), 
    \label{eq:VzH_expansion}
\end{equation}
Enforcing the no-slip BC $V_Z|_{Y=H}=0$ due to \eqref{eq:kinematic_bc_z}, \eqref{eq:VzH_expansion} requires that
\refstepcounter{equation}
$$
    V_{Z,0}|_{Y=1} = 0,\qquad
    V_{Z,1}|_{Y=1} = - \left. U_{Y,0} \frac{\partial V_{Z,0}}{\partial Y}\right|_{Y=1}.
    \eqno{(\theequation{\mathit{a},\mathit{b}})}
    \label{eq:Vz_BC_Y1}
$$%
Observe that the flow-induced deformation of the channel leads to an \emph{effective slip} velocity $V_{Z,1}|_{Y=1}$ in (\ref{eq:Vz_BC_Y1}\textit{b}) along the original location of the wall \citep{AC20,ZR24}.

Since we are interested only in the streaming (or rectified) flow component, we now apply the cycle-averaging operator $\langle \, \cdot \, \rangle \defeq \frac{1}{2\uppi} \int_{0}^{2\uppi} (\, \cdot \,) \,\rd T$ to \eqref{eq:zmom_3D_wide} and \eqref{eq:continuity_3D_2} at $O(\beta)$, to obtain
\begin{subequations}\label{eq:O2_avg_1}\begin{align}
    \frac{\Wor^2}{\gamma} \left\langle V_{Y,0} \frac{\partial V_{Z,0}}{\partial Y} + V_{Z,0} \frac{\partial V_{Z,0}}{\partial Z} \right\rangle &= - \frac{\rd \langle P_1\rangle }{\rd Z} + \frac{\partial^2 \langle V_{Z,1} \rangle}{\partial Y^2}, \label{eq:O2_avg_prob_p_control_mom}\\
    \frac{\partial \langle Q_1 \rangle}{\partial Z}  &= 0. 
\end{align}%
\end{subequations}
Observe that the left-hand side of \eqref{eq:O2_avg_prob_p_control_mom} is independent of $X$, and so is $\langle P_1 \rangle$ [due to the cycle-averaged \eqref{eq:xmom_3D_wide} at $O(\beta)$]. Thus, we can $X$-average the $O(\beta)$ problem's governing equations~\eqref{eq:O2_avg_1}:
\begin{subequations}\begin{align}
    \frac{\Wor^2}{\gamma} \left\langle V_{Y,0} \frac{\partial V_{Z,0}}{\partial Y} + V_{Z,0} \frac{\partial V_{Z,0}}{\partial Z} \right\rangle &= - \frac{\rd \langle P_1\rangle }{\rd Z} + \frac{\partial^2 \llangle V_{Z,1} \rrangle}{\partial Y^2}, \label{eq:O2_avg_prob_p_control_mom_2}\\
    \frac{\partial \llangle Q_1 \rrangle}{\partial Z}  &= 0. \label{eq:O2_avg_prob_p_control_cont_2}
\end{align}
\label{eq:O2_avg_prob_p_control}\end{subequations}%
where $\llangle \,\cdot\, \rrangle \defeq \frac{1}{2\uppi} \int_{-1/2}^{+1/2} \int_{0}^{2\uppi} (\, \cdot \,) \,\rd T\rd X$ denotes the simultaneous $T$ and $X$ averaging. The $T$-averages involving $O(1)$ phasors $\mathcal{A}=\Real[\mathcal{A}_a \re^{\ri T}]$ and $\mathcal{B}=\Real[\mathcal{B}_a \re^{\ri T}]$ are calculated by the standard rule $\langle \mathcal{A} \mathcal{B} \rangle = \tfrac{1}{2} \Real[\mathcal{A}_a^* \mathcal{B}_a] = \tfrac{1}{2} \Real[\mathcal{A}_a \mathcal{B}_a^*]$, where a star superscript denotes complex conjugate. 

Four conditions are required to simultaneously and uniquely determine $\llangle V_{Z,1}\rrangle$, $\llangle Q_1 \rrangle$, and $\langle P_1\rangle$ from \eqref{eq:O2_avg_prob_p_control}. The suitable BCs now correspond to no slip at $Y=0$ [from averaging (\ref{eq:no_slip_bc}\textit{a})] and effective slip at $Y=1$ [from averaging (\ref{eq:Vz_BC_Y1}\textit{b})]:
\refstepcounter{equation}
$$
    \llangle V_{Z,1} \rrangle|_{Y=0} = 0,\qquad 
    \llangle V_{Z,1} \rrangle |_{Y=1} = - \left\langle\!\left\langle \left. U_{Y,0} \frac{\partial V_{Z,0}}{\partial Y}\right|_{Y=1} \right\rangle\!\right\rangle.
    \eqno{(\theequation{\mathit{a},\mathit{b}})}
    \label{eq:O2_avg_V_bc_p_control}
$$
The conditions in the experiment are such that the membrane in the liquid-filled chamber used for oscillatory flow generation (recall figure~\ref{expstp}) does not allow any net flow through the system. Then, according to \eqref{eq:O2_avg_prob_p_control_cont_2}, $\llangle Q_1 \rrangle = \text{const.}$, and this constant must be zero throughout.  The outlet is open to the gauge pressure per \eqref{eq:pressure_bc}. Thus, the remaining BCs are
\refstepcounter{equation}
$$
    \llangle Q_1 \rrangle|_{Z=0} = 0,\qquad 
    \langle P_1 \rangle|_{Z=1} = 0. 
    \eqno{(\theequation{\mathit{a},\mathit{b}})}
    \label{eq:O2_avg_P_bc_p_control}
$$
Using \eqref{eq:beam_3D_soln}, the slip velocity becomes
\begin{equation}
    \left\langle\!\left\langle \left. U_{Y,0} \frac{\partial V_{Z,0}}{\partial Y}\right|_{Y=1} \right\rangle\!\right\rangle = \left\langle \int_{-1/2}^{+1/2} U_{Y,0} \,\rd X \left. \frac{\partial V_{Z,0}}{\partial Y}\right |_{Y=1} \right\rangle 
    = \left\langle \left. (1+\mathscr{T}) P_0 \frac{\partial V_{Z,0}}{\partial Y}\right |_{Y=1} \right\rangle .
    \label{eq:averaged_slip_vel}
\end{equation}

It is convenient to now rewrite $\llangle V_{Z,1}\rrangle$ in a way to eliminate the pressure gradient on the right-hand side of \eqref{eq:O2_avg_prob_p_control_mom_2} and simultaneously satisfy the slip BC (\ref{eq:O2_avg_V_bc_p_control}\textit{b}). Specifically, let
\begin{equation}
    \llangle V_{Z,1} \rrangle(Y,Z) = - \frac{1}{2}\frac{\rd \langle P_1\rangle }{\rd Z} Y(1-Y) 
    - Y (1+\mathscr{T}) \left.\left\langle P_0 \frac{\partial V_{Z,0}}{\partial Y} \right|_{Y=1}\right\rangle 
    + \widetilde{\llangle V_{Z,1} \rrangle}(Y,Z).  
    \label{eq:Vz1T_p_control}
\end{equation}
Then, substituing \eqref{eq:Vz1T_p_control} into  \eqref{eq:O2_avg_prob_p_control_mom_2}, we obtain a new BVP for $\widetilde{\llangle V_{Z,1} \rrangle}$, subject to homogeneous BCs:
\begin{equation}
\left. \begin{array}{l}
    \displaystyle
    \frac{\Wor^2}{\gamma} \left\langle V_{Y,0} \frac{\partial V_{Z,0}}{\partial Y} + V_{Z,0} \frac{\partial V_{Z,0}}{\partial Z} \right\rangle =  \frac{\partial^2 \widetilde{\llangle V_{Z,1} \rrangle}}{\partial Y^2},\\[16pt] 
    \displaystyle
    \widetilde{\llangle V_{Z,1} \rrangle}|_{Y=0} = \widetilde{\llangle V_{Z,1} \rrangle}|_{Y=1} = 0.
\end{array} \right\}
\label{eq:Vz1T_avg_prob}
\end{equation}
An analytical solution appears unlikely since the left-hand side of the ODE in \eqref{eq:Vz1T_avg_prob} is a complicated, complex-valued function of $Y$ and $Z$, but some useful approximations are discussed in Appendix~\ref{app:approximation}. Instead, we solve the linear two-point BVP~\eqref{eq:Vz1T_avg_prob} numerically using \textsc{Matlab}'s \texttt{bvp4c} subroutine, which implements a finite-difference solver with residual-based error control \citep{KS01}, with absolute and relative tolerance of $10^{-6}$.

Finally, from \eqref{eq:O2_avg_prob_p_control_cont_2}, (\ref{eq:O2_avg_P_bc_p_control}\textit{a}) and \eqref{eq:Vz1T_p_control}, we conclude that
\begin{equation}
    \llangle Q_1 \rrangle = \int_0^1 \llangle V_{Z,1} \rrangle \,\rd Y = - \frac{1}{12}\frac{\rd \langle P_1\rangle }{\rd Z} + \mathfrak{Q}(Z) = 0,
    \label{eq:avg_Q1}
\end{equation}
where, for convenience, we have let
\begin{equation}
    \mathfrak{Q}(Z) \defeq \underbrace{- \frac{1}{2} (1+\mathscr{T}) \left\langle \left. P_0 \frac{\partial V_{Z,0}}{\partial Y} \right|_{Y=1}\right\rangle}_{\text{from effective slip}} 
    + \underbrace{\int_0^1 \widetilde{\llangle V_{Z,1} \rrangle} \, \rd Y}_{\text{from advective inertia}}.
    \label{eq:funny_Q}
\end{equation}
Equation~\eqref{eq:avg_Q1} is a first-order differential equation for the streaming pressure $\langle P_1 \rangle$ subject to the outlet BC (\ref{eq:O2_avg_P_bc_p_control}\textit{b}), which is easily solved to obtain:
\begin{equation}
    \langle P_1 \rangle (Z) = - 12 \int_Z^1 \mathfrak{Q}(\tilde{Z})\,  \rd \tilde{Z}.
    \label{eq:avg_P1_Z}
\end{equation} 
In the experiment, the membrane in the liquid-filled chamber used for oscillatory flow generation imposes a weak, $O(\beta)$, nonzero mean pressure at the inlet, which is consistent with \eqref{eq:avg_P1_Z}.
To plot $\langle P_1 \rangle(Z)$, we evaluate $\mathfrak{Q}(Z)$ from \eqref{eq:funny_Q} wherein the integral over $Y$ is computed numerically using the trapezoidal rule via \textsc{Matlab}'s \texttt{trapz} with $\Delta Y=0.0101$ from the numerical solution for $\widetilde{\llangle V_{Z,1} \rrangle}$ of BVP~\eqref{eq:Vz1T_avg_prob}. Then, the indefinite integral in \eqref{eq:avg_P1_Z} is evaluated numerically using \textsc{Matlab}'s \texttt{cumtrapz} using $\Delta Z=0.0204$.


\section{Comparison between experiment and theory}
\label{sec:discussion}

To make the comparison between the experimental measurements (\S~\ref{sec:exp}) and the primary and streaming pressure distributions predicted by the theory (\S~\ref{sec:ei_recti}), we first post-processed the experimental data. To isolate the primary pressure oscillations and set the gauge pressure to zero, the mean of the signal was removed. Then, we fit the oscillatory pressure experimental data from the inlet pressure sensor 1 to a sinusoidal waveform of the form $p_1(t) = p_0\cos(2\uppi f_{\mathrm{true}}(t-t_{\mathrm{tshift}}))$, where $p_0$ is the amplitude of the pressure signal we seek to determine, $f_{\mathrm{true}}$ is the ``true'' frequency of the signal (slightly shifted from the input waveform frequency $f$ due to imperfections in the system) and $t_\mathrm{shift}$ is a phase introduced by the fact that the experimental data capture does not have to start right at a peak or trough of the sinusoidal signal. To find $p_0$, we applied \textsc{MATLAB}'s \texttt{findpeaks} subroutine to the zero-mean signal, and the values it returned were averaged to obtain $p_0$. Then, $f_{\mathrm{true}}$ and $t_{\mathrm{shift}}$ were found using \texttt{findfit}. After $p_0$ and $f_{\mathrm{true}}$ were successfully identified for a given experiment, the key dimensionless numbers (table \ref{tab:DimlessNum}) were calculated and used to evaluate the theoretical predictions.

\begin{figure}
    \centering
    \begin{subfigure}[b]{0.45\textwidth}
        \centering
        \caption{}
        \includegraphics[width=\textwidth]{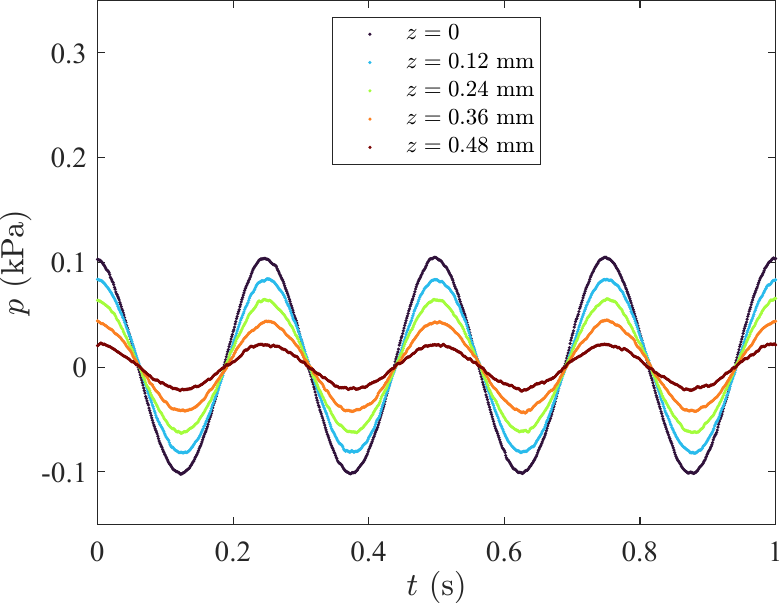}
    \end{subfigure}
    \quad
    \begin{subfigure}[b]{0.45\textwidth}
        \centering
        \caption{}
        \includegraphics[width=\textwidth]{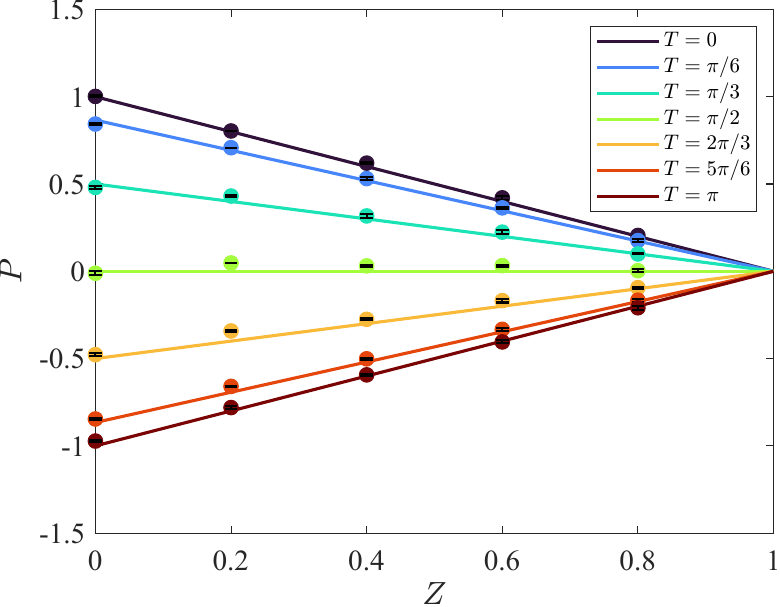}
    \end{subfigure}
    \caption{Pressure distribution and evolution in a rigid channel ($\gamma=0$) with $\Wor=2.5$. (\textit{a}) Experimental time series of the evolution of the pressure over time at the different axial positions of the ports, recall figure~\ref{expstp}(\textit{b}). (\textit{b}) Comparison between the evolution of the dimensionless axial pressure distribution from the experiments (solid symbols) and the rigid-channel theory (solid lines) over half a cycle.}
    \label{fig:p_rigid_channel}
\end{figure}

\subsection{Validation in a rigid channel}
\label{sec:compare_rigid}

To validate the experimental system, we first assessed its performance using a rigid channel ($\gamma=0$) with DI water as the working fluid. This validation is accomplished by comparing the experimental pressure data for the rigid channel to the $\gamma\to0$ limit of the theory, which gives $P(Z,T) = \Real[(1-Z)\re^{\ri T}]$ from \eqref{eq:Pa0_soln}. We considered three validation cases with different input frequencies (corresponding to $\Wor = 2.5$, $3.32$, and $3.96$). The case of $\Wor = 2.5$ is shown as an example in figure~\ref{fig:p_rigid_channel}. We observe good agreement between the theory and experiments for the axial distribution of $P$ and its variation over time in figure~\ref{fig:p_rigid_channel}(\textit{b}). The same holds for the other two validation experiments (not shown). 

Note that the experimental pressure data time-series [figure~\ref{fig:p_rigid_channel}(\textit{a})] does not show any phase difference between the pressure signals at the different axial positions, which is to be contrasted with the results for the deformable channel below (\S~\ref{sec:compare_deformable}). The signals' amplitudes exhibit a trend of linear attenuation with $z$, characterized by a constant multiplicative relationship between the values at different $z$. Correspondingly, in figure~\ref{fig:p_rigid_channel}(\textit{b}), these observations are reflected in the linear variation of $P$ with $Z$ at every $T$ over half an oscillation cycle.

\begin{figure}
    \centering
    \begin{subfigure}[b]{0.45\textwidth}
        \centering
        \caption{\hspace*{2em}$\Wor=0.537$, $\beta=0.0208$}
        \includegraphics[width=\textwidth]{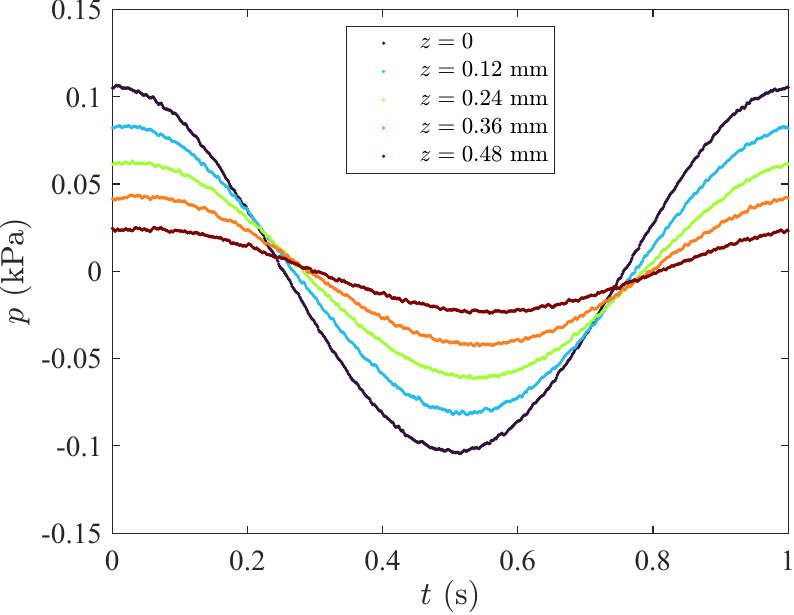}
    \end{subfigure}
    \quad
    \begin{subfigure}[b]{0.45\textwidth}
        \centering
        \caption{\hspace*{2.2em}$\Wor=0.537$, $\beta=0.164$}
        \includegraphics[width=\textwidth]{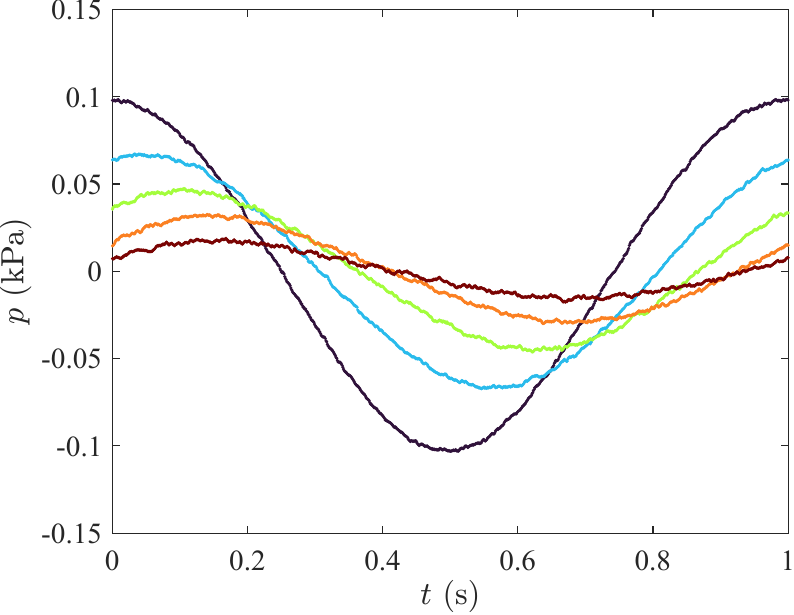}
    \end{subfigure}
    \begin{subfigure}[b]{0.45\textwidth}
        \centering
        \caption{\hspace*{2.2em}$\Wor=2.15$, $\beta=0.0167$}
        \includegraphics[width=\textwidth]{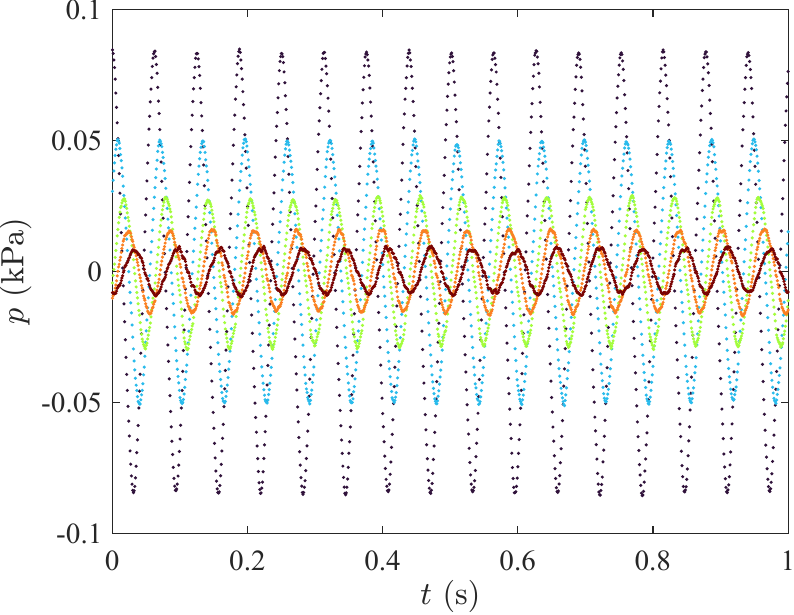}
    \end{subfigure}
    \quad
    \begin{subfigure}[b]{0.45\textwidth}
        \centering
        \caption{\hspace*{2.4em}$\Wor=2.15$, $\beta=0.104$}
        \includegraphics[width=\textwidth]{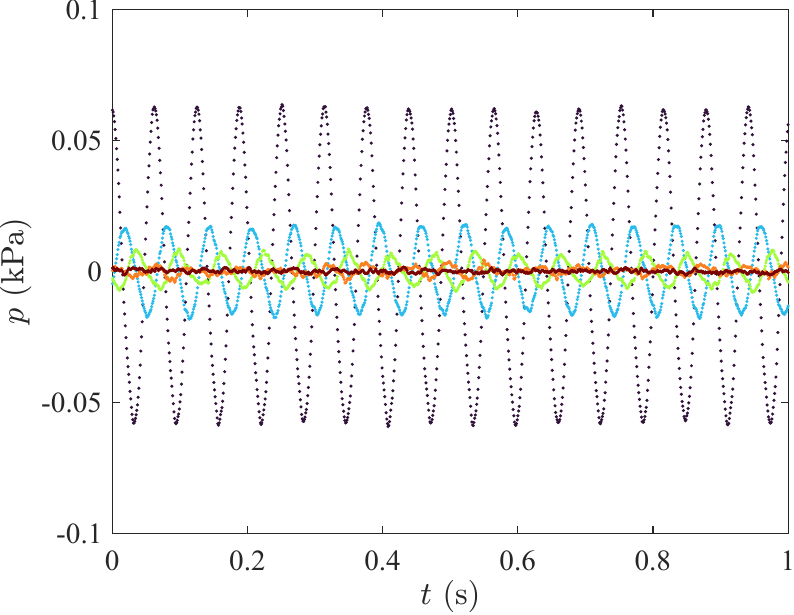}
    \end{subfigure}
    \caption{Experimental measurements of the evolution of the pressure over time at different axial positions (pressure port locations) for the deformable channel for smaller and larger compliance numbers (left column versus right column) and smaller and larger Womersley numbers (top row versus bottom row). Specifically, (\textit{a}) $\Wor=0.537$ ($\gamma=0.109$), (\textit{b}) $\Wor=0.537$ ($\gamma=0.913$), (\textit{c}) $\Wor=2.15$ ($\gamma=1.745$), and (\textit{d}) $\Wor=2.15$ ($\gamma=14.6$), respectively.}
    \label{fig:soft_expt}
\end{figure}

\subsection{Comparison of primary pressure oscillations in deformable channels}
\label{sec:compare_deformable}

Next, we turn to the experiments in the deformable channels, based on the 50 wt\% glycerin solution as the working fluid.  In this subsection, we discuss the \emph{primary} (purely oscillatory) pressure (\S~\ref{sec:primary_flow}) in the compliant channel. In figure~\ref{fig:soft_expt}, we show the experimentally measured pressure evolution at each of the different pressure ports for four pairs of values of the Womersley and compliance numbers---a low value $\Wor=0.537$ (slow flow oscillation) and a high value $\Wor=2.15$ (fast flow oscillation), as well as two different orders of magnitude of $\beta$, namely $\simeq 10^{-2}$ and $\simeq 10^{-1}$. To change the compliance number $\beta$, the thickness of the deformable top wall was varied in the experiments (recall table~\ref{tab:channel}). In figure~\ref{fig:soft_expt}, the mean pressure has been subtracted to set the outlet pressure as the gauge and to make the signal purely oscillatory.

Comparing the pressure time series in figure~\ref{fig:soft_expt} [for example, panel (\textit{b})] with time series in the rigid channel in figure ~\ref{fig:p_rigid_channel}(\textit{a}), we observe a distinct phase difference developing between the time series collected at different $z$ (\textit{i.e.}, at different pressure ports). Furthermore, the decrease in the signals' amplitudes is not proportional in the deformable channel, unlike the rigid channel. These differences are expected to arise from the nonlinear two-way coupling of the flow and deformation (increasing $\beta$).

\begin{figure}
    \centering
    \begin{subfigure}[b]{0.45\textwidth}
        \centering
        \caption{\hspace*{2em}$\Wor=0.537$, $\gamma=0.109$}
        \includegraphics[width=\textwidth]{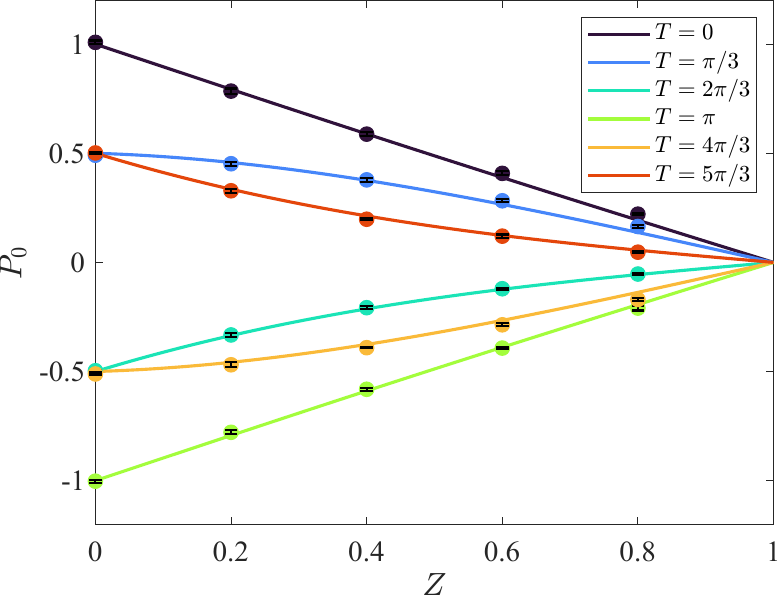}
    \end{subfigure}
    \quad
    \begin{subfigure}[b]{0.45\textwidth}
        \centering
        \caption{\hspace*{2em}$\Wor=0.537$, $\gamma=0.913$}
        \includegraphics[width=\textwidth]{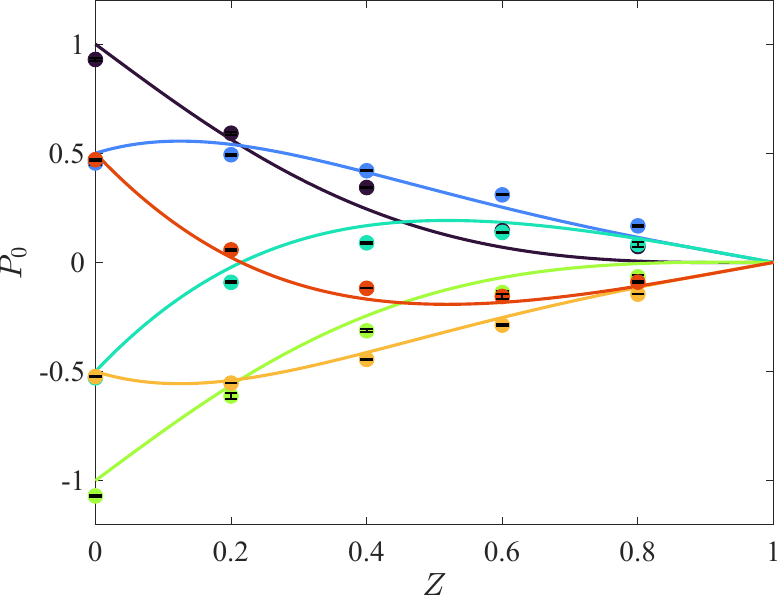}
    \end{subfigure}
    \begin{subfigure}[b]{0.45\textwidth}
        \centering
        \caption{\hspace*{2.3em}$\Wor=1.42$, $\gamma=0.763$ }
        \includegraphics[width=\textwidth]{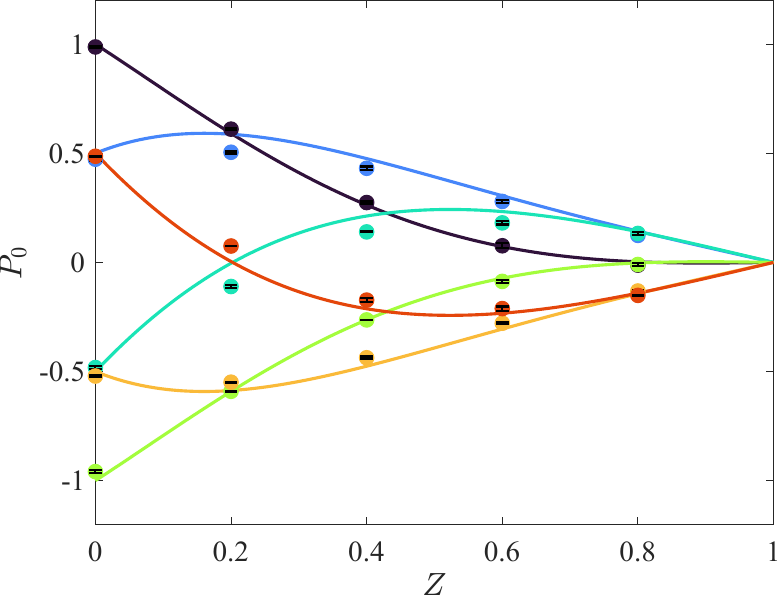}
    \end{subfigure}
    \quad
    \begin{subfigure}[b]{0.45\textwidth}
        \centering
        \caption{\hspace*{2.4em}$\Wor=1.42$, $\gamma=6.40$}
        \includegraphics[width=\textwidth]{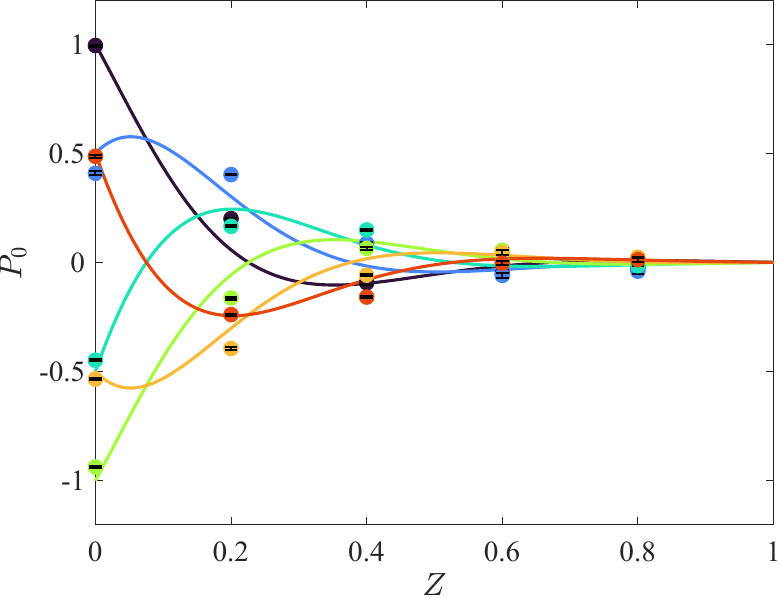}
    \end{subfigure}
    \begin{subfigure}[b]{0.45\textwidth}
        \centering
        \caption{\hspace*{2.5em}$\Wor=2.15$, $\gamma=1.745$}
        \includegraphics[width=\textwidth]{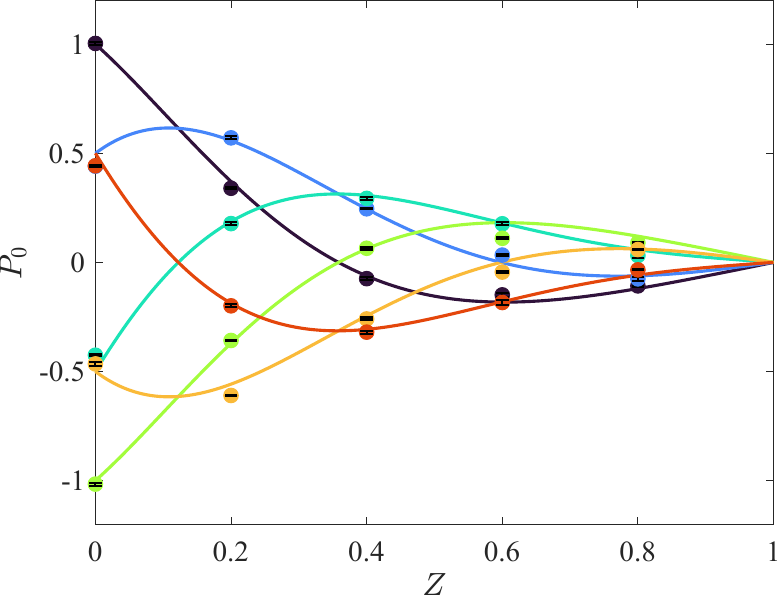}
    \end{subfigure}
    \quad
    \begin{subfigure}[b]{0.45\textwidth}
        \centering
        \caption{\hspace*{2.5em}$\Wor=2.15$, $\gamma=14.6$}
        \includegraphics[width=\textwidth]{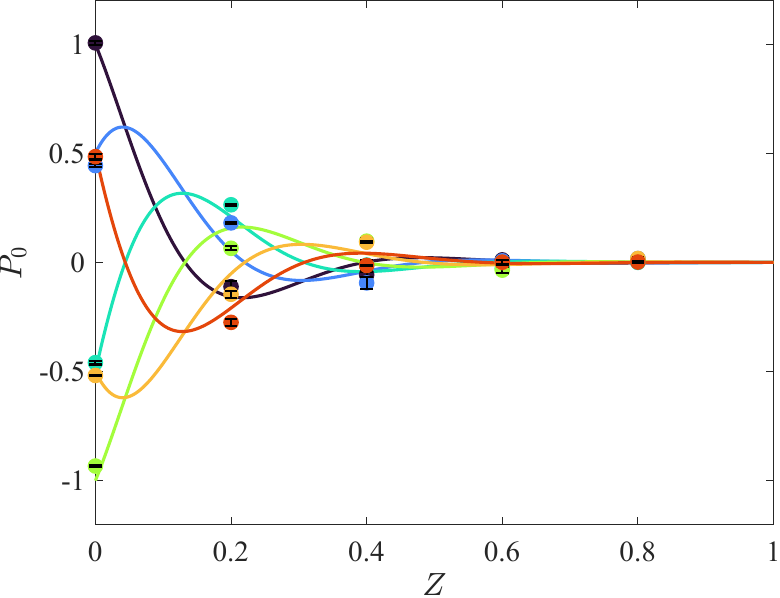}
    \end{subfigure}
    \caption{Comparison of the dimensionless axial pressure distribution in a deformable channel between experiment (solid symbols) and theory (solid curves), \textit{i.e.}, $P_0(Z,T) = \Real[P_{0,a}(Z)\re^{\ri T}]$ based on \eqref{eq:Pa0_soln}, with (\textit{a}) $\Wor=0.537$ ($\beta=0.0208$), (\textit{b}) $\Wor=0.537$ ($\beta=0.164$), (\textit{c}) $\Wor=1.42$ ($\beta=0.02$), (\textit{d}) $\Wor=1.42$ ($\beta=0.125$), (\textit{e}) $\Wor=2.15$ ($\beta=0.0167$), and (\textit{f}) $\Wor=2.15$ ($\beta=0.104$) respectively. The evolution of the pressure distribution is shown over a full cycle (thus $T=2\uppi$ overlaps $T=0$).}
    \label{fig:soft_comparison}
\end{figure}

To clearly demonstrate the nonlinear coupling, in figure~\ref{fig:soft_comparison}, we compare the dimensionless primary pressure distribution $P_0(Z,T) = \Real[P_{0,a}(Z)\re^{\ri T}]$ from \eqref{eq:Pa0_soln} predicted by the theory to the experimental data. We neither construct $P_1(Z,T)$, which is not straightforward \citep{ZR24}, nor neglect its $O(\beta)$ contribution, as we have ensured a one-to-one comparison of primary theoretical and experimental pressures by removing the mean of the experimental time series. Notably, unlike the steady \citep{CCSS17} and startup \citep{MCSPS19} problems, the key dimensionless groups influencing the pressure distribution in the deformable channel are the Womersley number $\Wor$ and the elastoviscous number $\gamma$, not the compliance number $\beta$.

The results in figure~\ref{fig:soft_comparison} show a good agreement between the theory of the primary pressure oscillations $P_0(Z,T)$ and the experimental measurements over a wide range of $\Wor$ and $\gamma$. We observe that the agreement between theory and experiment is better for $0.1<\gamma<1$ than for $\gamma>1$. This deviation can be attributed to the fact that for larger values of $\gamma>1$, the combined effect of compliance and oscillations is strong, and the time it takes the top wall to adjust to the flow oscillations becomes much longer than the oscillation timescale, localizing the majority of the pressure variation near the channel's inlet. This rapid, localized variation is more challenging to capture using equally spaced pressure ports in the experiments. Nevertheless, the overall agreement between theory and experiment on the trend of $P_0(Z,T)$ and how it changes with $\Wor$ and $\gamma$ is good, thus not only validating the predicted nonlinear pressure distribution~\eqref{eq:Pa0_soln} but also providing the first experimental demonstration of the strong coupling between flow oscillations and wall deformations, even in weakly compliant channels ($\beta\ll1$).


\subsection{Comparison of streaming pressure profiles in deformable channels}
\label{sec:compare_streaming}

\begin{figure}
     \centering
     \begin{subfigure}[b]{0.45\textwidth}
         \centering
         \caption{}
         \includegraphics[width=\textwidth]{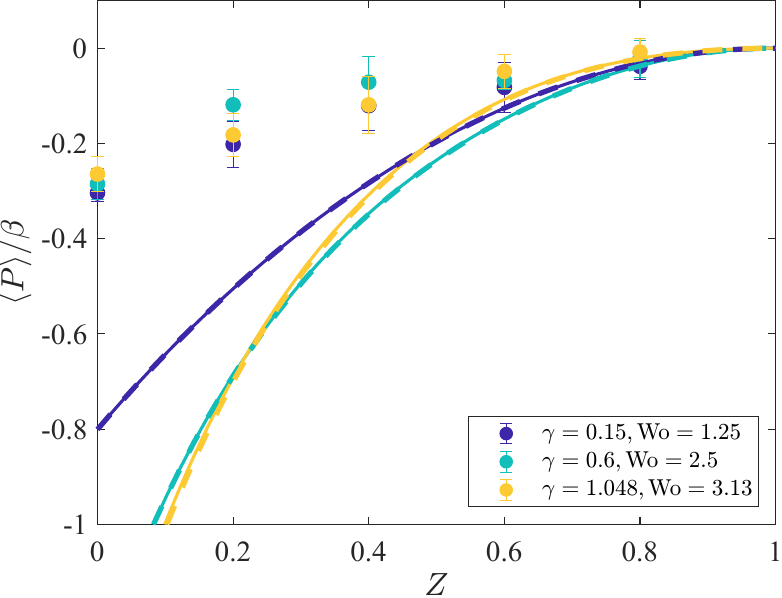}
     \end{subfigure}
     \quad
     \begin{subfigure}[b]{0.45\textwidth}
        \centering
        \caption{}
        \includegraphics[width=\textwidth]{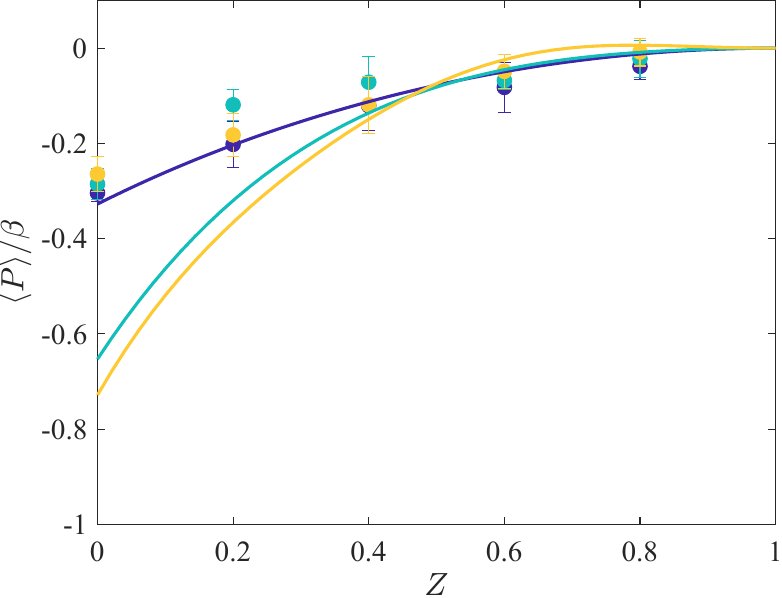}
     \end{subfigure}
     \caption{(\textit{a}) Comparison of streaming pressure (cycle-averaged pressure) distribution, $\langle P \rangle/\beta = \langle P_1 \rangle$, between the experimental data (solid symbols) the theoretical prediction based on numerically evaluating \eqref{eq:avg_P1_Z} (solid curves) and the closed-form approximation \eqref{eq:avg_P1_Z_approx} (dashed curves, overlapping the solid curves) in a deformable channel with $\beta\approx0.17$ and $\gamma=0.15$ ($\Wor = 1.25$), $\gamma=0.6$ ($\Wor = 2.5$), and $\gamma=1.048$ ($\Wor = 3.13$) achieved by changing the input frequency in the same channel. (\textit{b}) \textit{Ad hoc} modification of the theory by multiplying the effective slip term in \eqref{eq:funny_Q} by $0.4$ to demonstrate the sensitivity of the streaming pressure to the effective slip contribution.}
     \label{fig:streaming}
 \end{figure}

Next, we turn to the elastoinertial rectification phenomenon, namely the theoretical prediction that $\langle P \rangle/\beta \equiv \langle P_1 \rangle \ne 0$ due to the nonlinear coupling of the flow's inertia with the wall deformation (\S~\ref{sec:secondary_flow}). All data shown in this section is based on DI water as the working fluid. To this end, in figure~\ref{fig:streaming}, we compare the streaming pressure $\langle P_1 \rangle(Z)$ calculated numerically from \eqref{eq:avg_P1_Z} (as described above) to the corresponding quantity extracted from the experiments. 

This comparison is more challenging than the previous one in \S~\ref{sec:compare_deformable} because we are now dealing with small quantities that are $O(\beta)$. Consequently, the error bars on the experimental data in figure~\ref{fig:streaming} are much larger as the pressure values being measured push the experimental system to its sensitivity limit. Nevertheless, in figure~\ref{fig:streaming}(\textit{a}), we see a reasonable agreement between experiment and theory regarding the \emph{trend} of the streaming pressure distribution along the channel. Interestingly, $\langle P_1\rangle$ is less sensitive to $\gamma$ under the present flow conditions, and both the theory curves and experimental data cluster together. 

The largest disagreement is at $Z=0$, at the first pressure port, which may be expected as this is the location in the experimental system that is least likely to satisfy all the assumptions of the theory. Specifically, as shown in figure~\ref{expstp}, there is a rigid section attached to the inlet of the deformable channel in the experiments, which constrains the displacement along the inlet plane. However, this possibility is not accounted for in the theory---the leading-order equation~\eqref{eq:thick_plate_eqs_2} for the displacement does not allow for boundary conditions to be imposed in $Z$ and thus cannot capture this localized of clamping at the inlet, which previous work showed is confined to a ``boundary layer'' of thickness $O(\sqrt{w/\ell})$ \citep{WC21}. While the inlet conditions on the displacement have no discernible effect on $P_0(Z)$, as demonstrated in \S~\ref{sec:compare_deformable} by the excellent agreement between theory and experiments at $Z=0$, these conditions may affect the \emph{weaker} effects being investigated at $O(\beta)$.

To test the hypothesis that the inlet conditions may affect the agreement there, we turn to \eqref{eq:funny_Q} and \eqref{eq:avg_P1_Z}, from which we observe that the streaming pressure is generated by a competition between effective wall slip (at the location of the undeformed wall) and advective inertia. Specifically, the effective slip is a direct function of the displacement, per \eqref{eq:averaged_slip_vel}. Therefore, we expect this $O(\beta)$ quantity to be possibly strongly affected by the inlet restrictions in the experiments. Specifically, if the displacement in the experiments near the inlet is constrained, or otherwise reduced, then this term might be overestimated by the theory. To test this hypothesis, we check the sensitivity of the $\langle P_1\rangle(Z)$ profile to the magnitude of the effective slip term. We find that even approximately halving this term can account for a lot of the disagreement between theory and experiment, especially as $Z\to0$, as shown by the \textit{ad hoc} modification in figure~\ref{fig:streaming}(\textit{b}). 

Though the \textit{ad hoc} modification improves the agreement quite a bit for the lowest frequency (lowest $\Wor$ and $\gamma$ values), it still exhibits disagreement for the larger frequencies (larger $\Wor$ and $\gamma$ values). In fact, fully suppressing the effective slip term brings the streaming pressures at the higher frequency even closer to the experimental data points. To further investigate the frequency-dependent nature of this agreement/disagreement, we would need to better assess how well our theory models the effective slip term in \eqref{eq:funny_Q}. The present theory evaluates this term using the result that $U_{Y,0} \propto P_0$, per \eqref{eq:averaged_slip_vel}, and the shear $\partial V_{Z,0}/\partial Y$ based on \eqref{eq:Vza0_soln}. We would need a more advanced experimental setup, which measures the velocity field and the deformation independently, to evaluate the accuracy of each of these results. Alternatively, measuring just the slip velocity at $Y=1$ experimentally, say using particle image velocimetry (PIV), and comparing it to (\ref{eq:Vz_BC_Y1}\textit{b}) could be another independent check. These directions will be pursued in forthcoming investigations.

Despite the limitations and challenges of these measurements, the experimental data appear to capture the key effects of $\Wor$ and $\gamma$ on the streaming pressure profile, including the nonmonotonic behavior with respect to $\gamma$ (in particular for $Z>0.5$), though admittedly, the error bars on the experimental measurements for different $\gamma$ overlap.

\section{Conclusion}
\label{sec:conclusion}

We presented a systematic, combined theoretical and experimental investigation of two-way coupling between oscillatory internal viscous flows and deformable confining boundaries. Specifically, we provided a theory (and solutions) for the pressure distribution due to oscillatory flow in a 3D channel with a deformable top wall, its relation to the flow rate in terms of complex-valued functions, as well as the shape of the deformation of the compliant wall, taking into account its thickness. Consistent with microfluidics-oriented applications, we assumed the channels were shallow and slender, which allowed the use of the lubrication approximation. However, as convective inertia cannot be eliminated from the axial momentum equation, we also assumed a small compliance number to make progress on the nonlinearly coupled problem, unlike steady \citep{CCSS17} and start-up \citep{MCSPS19} flow-induced deformation problems previously analyzed. However, we did not make assumptions on the two key dimensionless groups involving the oscillation frequency: the Womersley number and the elastoviscous number, the latter being the key controlling parameter of this type of ``viscous--elastic'' structure interaction problem \citep{EG14}. To validate the theory, we designed a PDMS-based microfluidic experimental platform capable of measuring the pressure distribution in these flows. 

Our key findings are that the primary (periodic) pressure distribution from the theory shows strong agreement with the experimental measurements. Furthermore, we were able to measure (albeit with higher uncertainty) the weak, secondary (cycle-averaged, streaming) pressure distribution predicted by the theory. The resulting comparison shows agreement in the trends and thus provides the first experimental demonstration of \emph{elastoinertial rectification} due to oscillatory flow in 3D deformable channels, which is a subtle effect not previously measured in experiments. Our theoretical--experimental results demonstrate that, as \citet{ZR24} recently clarified, geometric nonlinearity due to the deformation of the channel and inertial nonlinearity due to the advective inertia of the fluid, are inextricably coupled in determining the pressure characteristics of these flows. Our work thus advances the fundamental understanding of soft hydraulics involving oscillatory flows. Notably, there are no fitting parameters in the theory; each property of the fluid, the deformable wall, and the geometry was experimentally characterized. Consequently, the theory of elastoinertial rectification in 3D deformable channels is ready for use in applications. 

In the future, it would be worth pursuing the experimental measurement of the phasing between the flow rate and the primary pressure (gradient), recently explored in tubes through simulations by \citet{KB25}, as well as the direct measurement of the elastic wall's deformation profile. Although our experiments showed negligible effect of the deformable wall's inertia, solving for the displacement profile for finite $\Str_s$ in \eqref{eq:thick_plate_eqs_2} would be relevant for applications to soft robotics \citep{BSBGO18}. In addition, many relevant working fluids for applications, such as polymer solutions, colloidal suspensions, and biological fluids, show non-Newtonian rheology \citep{CR08book,roselli2011bio}, which will introduce another set of nonlinear couplings beyond those already understood in steady flow \citep{C21,CBCF24,BC22}, between the fluid rheology (\textit{e.g.}, viscoelastic stresses or changes of the apparent viscosity due to shear-thinning), flow oscillations, and wall deformation. Further investigation of oscillatory flows of complex fluid in deformable channels, especially viscoelastic ones \citep{asghari2020oscillatory}, will be relevant to microfluidic-oriented applications \citep{DDS20,Muduetal24}. Another avenue of future work could be to revisit the possibility of flow rectification due to oscillatory flow in deformable poroelastic media \citep{FPM23}.



\backsection[Acknowledgements]{We thank Bhargav Rallabandi for many insightful discussions on elastoinertial rectification and Jon Coonley for experimental assistance.}

\backsection[Funding]{J.F.\ acknowledges partial support by the U.S.\ National Science Foundation under Grant No.\ CBET-2323045. I.C.C.\ acknowledges partial support by the U.S.\ National Science Foundation under Grant No.\ CMMI-2245343.}

\backsection[Declaration of interests]{The authors report no conflict of interest.}


\backsection[Author ORCIDs]{\\
\includegraphics[width=6px]{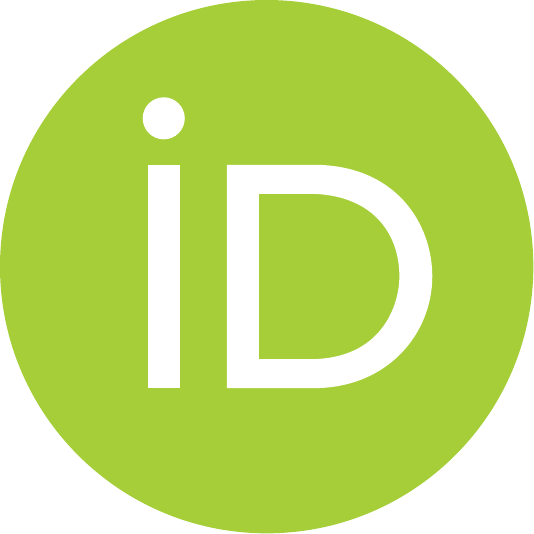} Anxu Huang, {\urlstyle{same}\url{https://orcid.org/0009-0001-8533-3343}};\\
\includegraphics[width=6px]{ORCID.pdf} Shrihari D.\ Pande, {\urlstyle{same}\url{https://orcid.org/0000-0001-6962-8400}};\\
\includegraphics[width=6px]{ORCID.pdf} Jie Feng, {\urlstyle{same}\url{https://orcid.org/0000-0002-4891-9214}};\\
\includegraphics[width=6px]{ORCID.pdf} Ivan C.\ Christov, {\urlstyle{same}\url{http://orcid.org/0000-0001-8531-0531}}.}




\appendix

\section{Closed-form approximations for the streaming quantities, and limits}
\label{app:approximation}

In \S~\ref{sec:ei_recti}, we solved the linear BVP~\eqref{eq:Vz1T_avg_prob} numerically since the left-hand side of the ODE is an unwieldy expression involving complex-valued quantities. \citet{ZR24} used an approximation procedure, similar to the one used to derive reduced models of cardiovascular flows \citep[see, \textit{e.g.},][]{vosse2011pulse}, that leads to closed-form, albeit \textit{ad hoc} (unless $\Wor\ll1$), expressions. The procedure consists of replacing the Womersley profile~\eqref{eq:Vza0_soln} with a Poiseuille profile \emph{with the same centerline velocity}. In the present context, the approximation takes the form:
\begin{equation}
    \left. \begin{array}{l}
    \displaystyle
    {V}_{Z,0,a}(Y,Z) \approx 
    4 \mathfrak{v}(\Wor) Y(1-Y) \left( -\frac{\rd P_{0,a}}{\rd Z} \right),\\[16pt]
    \displaystyle
    \mathfrak{v}(\Wor) \defeq \frac{1}{\ri{\Wor}^2}\left[1-\frac{1}{\cos\left(\ri^{3/2} {\Wor}/2 \right)}\right].
\end{array}\right\}
\label{eq:Vz0_approx}
\end{equation}
Substituting \eqref{eq:Vz0_approx} into \eqref{eq:com_3D_wide}, we find the approximate vertical velocity component:
\begin{equation}
     {V}_{Y,0,a}(Y,Z) \approx \mathfrak{v}(\Wor)\left(2Y^2-\frac{4Y^3}{3}\right)\kappa(\Wor,\gamma,\mathscr{T})^2 P_{0,a}(Z),
     \label{eq:Vy0_approx}
\end{equation}
having used \eqref{eq:P0_BVP} to replace ${\rd^2 P_{0,a}}/{\rd Z^2}$ by $\kappa^2 P_{0,a}$.
Using \eqref{eq:Vz0_approx} and \eqref{eq:Vy0_approx}, and carefully tracking the conjugation in evaluating the time-averages of phasors, we find an approximate solution of the BVP~\eqref{eq:Vz1T_avg_prob}:
\begin{equation}
    \widetilde{\llangle V_{Z,1}\rrangle} (Y,Z) \approx \frac{\Wor^2}{\gamma} \frac{1}{2} \Real\left[ |\mathfrak{v}|^2 \frac{\rd P_{0,a}}{\rd Z} (\kappa^2)^* P_{0,a}^* \right] \frac{2}{45} (4Y^6 - 12Y^5 + 15Y^4 - 7Y).
   \label{eq:Vz1T_avg_approx}
\end{equation}

Note that the approximate velocity profile \eqref{eq:Vz0_approx}--\eqref{eq:Vy0_approx} is used only for evaluating the \emph{advective} terms to obtain the closed-form solution \eqref{eq:Vz1T_avg_approx}. The slip velocity~\eqref{eq:averaged_slip_vel} can be calculated without approximation by using \eqref{eq:Vza0_soln} in \eqref{eq:averaged_slip_vel}, to find:
\begin{equation}
    \left\langle \left. P_0 \frac{\partial V_{Z,0}}{\partial Y}\right |_{Y=1} \right\rangle
    = \frac{1}{2}\Real\left[ P_{0,a}^*\frac{\ri^{1/2}}{\Wor}\tan\big(\ri^{3/2} \Wor/2\big)\left(-\frac{\rd P_{0,a}}{\rd Z} \right) \right].
    \label{eq:avg_P0dVz0dY_approx}
\end{equation}

Now, substituting \eqref{eq:Vz1T_avg_approx} and \eqref{eq:avg_P0dVz0dY_approx} into \eqref{eq:funny_Q}, we obtain
\begin{equation}
    \mathfrak{Q}(Z) \approx -(1 + \mathscr{T}) \Real\left[ \mathfrak{q}(\Wor) \frac{\rd P_{0,a}}{\rd Z} P_{0,a}^*  \right],
    \label{eq:funny_Q_approx}
\end{equation}
where
\begin{equation}
    \begin{aligned}
    \mathfrak{q}(\Wor) &\defeq -\left[\frac{\ri^{1/2}}{4\Wor}\tan\big(\ri^{3/2} \Wor/2\big)  + \frac{3\Wor^2}{70}  \ri\frac{|\mathfrak{v}(\Wor)|^2}{\mathfrak{f}(\Wor)^*} \right] \\[8pt]
    &=  
    \begin{cases}
        \frac{1}{8} - \ri \frac{31}{1680} \Wor^2 -  \frac{31}{16800} \Wor^4 + O(\Wor^6), &\Wor \to 0 ,\\[8pt]
        \frac{3}{70} + \frac{(35\ri + 12) \ri^{1/2}}{140} \Wor^{-1} + O(\Wor^{-2}), &\Wor \to \infty.
    \end{cases}
    \end{aligned}
\end{equation}
Recall that $\mathfrak{v}(\Wor)$ is defined in \eqref{eq:Vz0_approx}, and $\mathfrak{f}(\Wor)$ is defined \eqref{eq:Qa0_soln}. 
Notice that $\mathfrak{Q}$ depends on $\gamma$ only through $P_{0,a}$. Within $\mathfrak{q}$, the first term in the parentheses arises from effective slip at the original location of the deformable wall (no approximation), while the second term is the contribution of advective inertia (approximated based on \eqref{eq:Vz0_approx}--\eqref{eq:Vy0_approx}).

Based on \eqref{eq:funny_Q_approx}, we compute $\int_Z^1 \mathfrak{Q}(\tilde{Z})\, \rd \tilde{Z}$ and \eqref{eq:avg_P1_Z} becomes
\begin{multline}
    \langle P_1 \rangle (Z)  
    \approx -12(1 + \mathscr{T}) \\
    \times \Real\left\{ \mathfrak{q}(\Wor) \frac{\kappa}{2|\sinh\kappa|^2} \left[ \frac{\sinh^2\big((1-Z) \Real[\kappa]\big)}{\Real[\kappa]}-\ri\frac{\sin^2\big((1-Z) \Imag[\kappa]\big)}{\Imag[\kappa]} \right] \right\}.
    \label{eq:avg_P1_Z_approx}
\end{multline}
The comparisons in figure~\ref{fig:streaming}(\textit{a}) above show that this \textit{ad hoc} approximation is actually extremely accurate across a range of $\gamma$ and $\Wor$ values.

In the ``quasi-rigid limit'' \citep{ZR24}, $\gamma\to0$, a simpler expression can be obtained since $\kappa,\Real[\kappa],\Imag[\kappa]\sim\sqrt{\gamma}\to0$, namely 
\begin{equation}
    \langle P_1 \rangle (Z) \approx -6 \left(1+\mathscr{T}\right)\Real[\mathfrak{q}(\Wor)]  (1-Z)^2 ,\qquad \gamma \to 0.
    \label{eq:avg_P1_approx}
\end{equation}



\bibliography{Ref,references}


\end{document}